\documentclass[12pt]{article}
\usepackage{titlesec}
\usepackage[margin = 1.0in]{geometry}
\usepackage{amsmath, amsthm}
 \usepackage{ragged2e}

\usepackage{soul,color}
\usepackage[table]{xcolor} 

\usepackage{textcomp}
\usepackage{gensymb}
\usepackage{calligra}
\usepackage{commath}

\usepackage{makecell}

\newcolumntype{L}{>{\centering\arraybackslash}m{4.6cm}}

\usepackage[font=footnotesize,labelfont=bf]{caption}

\usepackage{qtree}

\usepackage{pgf}
\usepackage{tikz}
\usetikzlibrary{arrows,automata}
\usetikzlibrary{shapes.geometric, arrows}

\usepackage{amssymb}
\tikzstyle{startstop} = [rectangle, rounded corners, minimum width=3cm, minimum height=1cm,text centered, draw=black, fill=red!30]

\tikzstyle{io} = [trapezium, trapezium left angle=70, trapezium right angle=110, minimum width=3cm, minimum height=1cm, text centered, draw=black, fill=blue!30]

\tikzstyle{process} = [rectangle, minimum width=3cm, minimum height=1cm, text centered, draw=black, fill=orange!30]

\tikzstyle{rect01} = [rectangle, minimum width=1cm, minimum height=0.5cm, text centered, draw=black, fill=white!30]

\tikzstyle{decision} = [diamond, minimum width=3cm, minimum height=1cm, text centered, draw=black, fill=green!30]

\tikzstyle{arrow} = [thick,->,>=stealth]
\tikzstyle{arrow2} = [thick,->]

\tikzset{circ/.style={draw,circle,node distance=2cm,inner sep=0.5pt},topath/.style={to path={|-(\tikztotarget)}}}

\usetikzlibrary{calc}
\usepackage{relsize}
\usepackage[final]{pdfpages}
\usepackage{graphicx}

\tikzstyle{startstop} = [rectangle, rounded corners, minimum width=3cm, minimum height=1cm,text centered, draw=black, fill=red!30]

\tikzstyle{io} = [trapezium, trapezium left angle=70, trapezium right angle=110, minimum width=3cm, minimum height=1cm, text centered, draw=black, fill=blue!30]

\tikzstyle{process} = [rectangle, minimum width=3cm, minimum height=1cm, text centered, draw=black, fill=orange!30]

\tikzstyle{rect01} = [rectangle, minimum width=1cm, minimum height=0.5cm, text centered, draw=black, fill=white!30]

\tikzstyle{decision} = [diamond, minimum width=3cm, minimum height=1cm, text centered, draw=black, fill=green!30]

\tikzstyle{arrow} = [thick,->,>=stealth]

\tikzset{circ/.style={draw,circle,node distance=2cm,inner sep=0.5pt},topath/.style={to path={|-(\tikztotarget)}}}

\usepackage{tikz-cd}
\usetikzlibrary{fit}

\usepackage[utf8]{inputenc}
\usepackage[T1]{fontenc}
\usepackage[english]{babel}
\usepackage[T1]{fontenc}

\usepackage{lmodern}
\usepackage{amsmath}
\usepackage{amsthm}
\usepackage{amsfonts}

\usepackage{tikz}

\usepackage{xfp}

\newcommand\x{1.5}
\newcommand\y{3}
\newcommand\z{1}

\tikzstyle{state}=[shape=circle,draw=blue!50,fill=darkgray!20] 
\tikzstyle{observation}=[shape=rectangle,draw=orange!50,fill=orange!20]
\tikzstyle{lightedge}=[<-, dashed]
\tikzstyle{mainstate}=[state,very thick]
\tikzstyle{mainedge}=[<-, thick]

\tikzstyle{rect01} = [rectangle, minimum width=0.2cm, minimum height=0.2cm, text centered, draw=black, fill=white!30]

\usepackage{booktabs, multirow} 
\usepackage{soul}

\usepackage{changepage,threeparttable} 

\usepackage[utf8]{inputenc}
\usepackage{pgf}
\usepackage{tikz}
\usetikzlibrary{positioning}
\usetikzlibrary{arrows,automata}

\usepackage{lmodern}
\usepackage{amsmath}
\usepackage{amsthm}
\usepackage{amsfonts}

\usepackage{authblk}

\usepackage{blindtext}

\theoremstyle{definition} 
\newtheorem{definition}{Definition}[section]

\theoremstyle{definition}
\newtheorem{theorem}{Theorem}[section]

\theoremstyle{definition}
\newtheorem{exmp}{Example}[section]

\theoremstyle{definition} 
\newtheorem{Algorithm}{Algorithm}[section]

\titleformat{\section}
  {\normalfont\fontsize{15}{15}\bfseries}{\thesection}{1em}{}
  \usepackage{parskip}
\usepackage{multicol}
\usepackage{blindtext}
\title{Applications of Convolutional Codes to DNA Codes and Error-Correction}
\author[1]{Paridhi Latawa} 
\author[2]{Nuh Aydin}

\affil[1]{Liberal Arts and Science Academy (LASA), Austin, TX, USA}
\affil[2]{Department of Mathematics, Kenyon College}
\date{} 

\begin{document}

\maketitle


\tableofcontents



\pagebreak 

\section{Abstract}

Convolutional codes are error-correcting linear codes that utilize shift registers to encode. These codes have an arbitrary block size and they can incorporate both past and current information bits. DNA codes represent DNA sequences and are defined as sets of words comprised of the alphabet ${A, C, T, G}$ satisfying certain mathematical bounds and constraints. The application of convolutional code models to DNA codes is a growing field of biocomputation. As opposed to block codes, convolutional codes factor in nearby information bits, which makes them an optimal model for representing biological phenomena. This study explores the properties of both convolutional codes and DNA codes, as well as how convolutional codes are applied to DNA codes. It also proposes revisions to improve a current convolutional code model for DNA sequences. 

\section{Introduction}


Convolutional codes are error-correcting codes based on shift registers for polynomial encoding and decoding. Unlike block codes, which operate on fixed blocks of bits with a predetermined size, convolutional codes have an arbitrary block size \cite{MIT Lecture 8}. Since their introduction in 1955 by Peter Elias (\cite{Elias}), convolutional codes have grown in popularity. 

DNA codes store retrievable genetic information in a strand comprised of the four nitrogenous bases of DNA: Adenine (A), Thymine (T), Cytosine (C), and Guanine (G). DNA codes are represented by a four-letter alphabet, A, T, C, G, which can be further represented in binary. There are various constraints DNA codes have to satisfy, as explained in Section 4. 

As DNA goes through the central dogma of genetics, it undergoes the processes of transcription and translation. Transcription is the conversion of DNA to mRNA sequences. In mRNA, Uracil (U) replaces Thymine (T), so the four possible nucleotide bases are A, C, U, and G. At this point, the mRNA sequence is grouped as codons, blocks of size 3. There are 64 possible groups of codons that can be formed from the 4 bases ${A, C, U, G}$. Translation is the encoding of mRNA sequences to amino acids and their respective proteins. 

The four nucleotide bases of DNA, A, C, T, G, represent a maximum of $\log_2(4) = 2$ bits of information \cite{Bashford}. The genetic code differentiates between pyrimidines (Thymine, Uracil, and Cytosine) and purines (Adenine or Guanine) bases. The difference between pyrimidines and purines is the number of heterocyclic rings in their respective bases. Transitions, which are mutations that preserve the pyrimidine or purine classification, are less damaging than transversions, mutations that do not preserve the respective classifications \cite{Wilhelm}. 

DNA computing was first introduced by Tom Head in 1987 and first successfully executed by Leonard Adleman in 1994 \cite{Limbachiya}. In his seminal paper \cite{Adleman}, Adleman demonstrated the feasibility of conducting computations at the molecular level. Adleman first utilized DNA codes to solve the Hamiltonian path problem, a form of the  traveling salesman problem \cite{Adleman}. Through this experiment, he demonstrated the potential for an intersection between the fields of mathematics, technology, and the life sciences. 

Convolutional codes are notably similar to DNA codes as they produce encoded sequences that take into account present and past bits of information. Since initial experiments conducted by May et al. in the early 2000s that explored the role of convolutional codes in biocomputation, their relevance has grown as they can take genetic processes into account \cite{May-cc}. While current convolutional code models have taken necessary requirements and characteristics of DNA codes into consideration, various issues still need to be addressed \cite{Liu}. Understanding how outside influences and biological characteristics can further factor into DNA encoding and its mathematical representation is relevant. 

This study first explores the characteristics of convolutional codes, such as their encoding and decoding processes. Then, it consolidates the properties and constraints for DNA codes. Subsequently, the relevance of convolutional codes to DNA codes is studied. Various current convolutional code models for DNA encoding are analyzed, and improvements to a current convolutional code model for DNA sequences are proposed.  

\section{Convolutional Codes}


Convolutional codes have been used, in conjunction with Reed-Solomon codes, by the European Space Agency and NASA for communication during space missions \cite{Hankerson}. Convolutional codes are linear codes that utilize shift registers  to produce output sequences from input sequences. Our treatment of convolutional codes is largely based on Chapter 8 of \cite{Hankerson}.

\subsection{Encoding of Binary Convolutional Codes}

An $(n, k, m)$ binary convolutional code produces $n$ output bits for a set of $k$ information bits and has generator polynomials of degree $\leq m$ \cite{Hankerson}. The special case of $k = 1 $ indicates that one message digit is inputted in the shift register before the next code digit is calculated. An $(n, k, m)$ convolutional code has the rate $k/n$,  meaning that $k$ information bits are inputted into a shift register at a time and $n$ bits in a codeword are computed \cite{Hankerson}. Let $\mathbb{Z}_2=\{ 0,1\}$ be the binary field. A convolutional code is mathematically defined as follows.


\begin{definition}\cite{Hankerson}
An $(n, k, m)$ binary convolutional code with generators $g_1(x),\dots, g_n(x)$, where $g_i(x) \in \mathbb{Z}_2[x]$ with $\deg(g_i(x))\leq m$, is the code consisting of all codewords $c(x) = (c_1(x), c_2(x), ... , c_n(x))$ where $c_i(x) = m(x)g_i(x)$, and $m(x) = m_0 + m_1x + m_2x^2 + ...$ is a polynomial of arbitrary degree in $\mathbb{Z}_2[x]$.
\end{definition}

Assuming $L$ is the constraint length (encoder memory), when a convolutional code receives an input of $k$ bits, the convolutional encoder produces an output of $n$ bits (where $n > k$) that is associated with the present $k$ bits and the previous $L - 1$ group(s) of inputted $k$ bits \cite{McGraw-Hill}. 
The memory array of a convolutional encoder consists of memory cells with one output that can be linearly combined where the coefficients of the combination determine the output. For binary codes, having a coefficient of 1 denotes that the output of a memory cell is used in the linear combination, while a coefficient of 0 denotes that the output is not used. When brought together, the coefficients form a generator matrix. 

Like cyclic codes, convolutional codes can easily be encoded and decoded using shift registers. In general, an $s$-stage shift register consists of $s$ registers and has an output that is a linear combination of the contents of the registers. Take a polynomial $a(x)$ to represent the input sequence inputted into a shift register, a polynomial $c(x)$ to represent the output sequence of a shift register, and define a generator polynomial $g(x)$ as the function that represents the action of the shift register to take the input sequence and produce the respective output sequence. The relation between the input polynomial, generator polynomial, and output polynomial can be represented as $c(x) = a(x)g(x)$. Polynomial multiplication is implemented via shift registers to encode cyclic codes. Additionally, polynomial division can be implemented by a feedback shift register (FSR) to decode linear cyclic codes \cite{Hankerson}. 

Encoding can occur in two ways: utilizing the constructed shift register to graphically obtain the output sequence or by computing the output polynomial $c(x)$. The  example below demonstrates both methods of encoding. 

\begin{exmp}
Assuming $g(x) = 1 + x^2$ and $a(x) = 1 + x$, a shift register can be constructed to compute $a(x)g(x) = c(x)$. $a(x)g(x)$ can also be computed directly.

We assume that the input polynomial $a(x) = 1 + x$ represents the input sequence $1100000$. 

Assume the initial state of the three registers, given by the polynomial $g(x) = 1 + x^2$, is 000. The  register below and the table that follows are used to obtain the output sequence. 

\vspace{-0.3 cm}

\begin{figure}[h!]
\begin{center}
\begin{tikzpicture}[node distance=2cm]
\node (s0)[rect01]{\large s$_0$};
\node (s1)[rect01,right of=s0, xshift=0.1cm]{\large s$_1$};
\node (s2)[rect01,right of=s1, xshift=0.1cm]{\large s$_2$};
\node (s3)[rect01,right of=s2, xshift=0.1cm]{\large s$_3$};

\draw [arrow] (s0) -- (s1);
\draw [arrow] (s1) -- (s2);
\draw [arrow] (s2) -- (s3);

\node (E) [circ,above of= s1, yshift=1cm, xshift = 0 cm]{+};
\draw[arrow] (s0)--(E);
\draw[arrow] (s2)--(E);

\node (G) [above of= E,node distance=1.25cm]{G};

\draw [thick,-stealth](E) -- (G);


\end{tikzpicture}
\caption [font=\small] {Shift register for the convolutional code with generator polynomial $g(x) = 1 + x^2$ and input sequence $a(x) = 1 + x$. The 4 square nodes, $S_0, S_1, S_2, S_3$ represent the 4 states of the shift register. The binary addition operator links states $S_0$ and $S_2$, which are the states that have a coefficient of 1 in the linear combination used to form $g(x)$. The states $S_1$ and $S_3$ have a coefficient of 0 in the linear combination and thus are not included in $g(x)$ and do not have directed arrows pointing towards the binary operator. }
\end{center}
\end{figure}
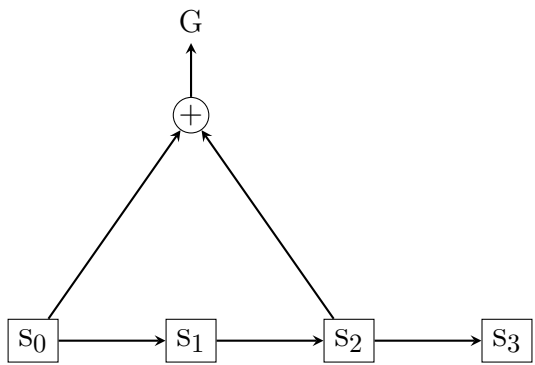

\vspace{-0.3 cm}

\begin{center}
\begin{table}[!htp]\centering
\caption{Table that shows how the output is obtained for the shift register in Figure 1. At time $t = -1$, the states $X_0, X_1, X_2$ contain 000. At $t = 0$, the first message digit is inputted into the shift register, as represented in the table.}
 \begin{tabular}{|c | c | c c c | c |} 
 \hline
Time & Input & $X_0$ & $X_1$ & $X_2$ & $Output = X_0 + X_2$ \\ [0.5ex]
\hline
 
 -1 & --- & 0 & 0 & 0 & --- \\
 0 & 1 & 1 & 0 & 0 & 1  \\
 1 & 1 & 1 & 1 & 0 & 1 \\
 2 & 0 & 0 & 1 & 1 & 1\\
3 & 0 & 0 & 0 & 1 & 1 \\
4 & 0 & 0 & 0 & 0 & 0 \\
5 & 0 & 0 & 0 & 0 & 0 \\  [1ex]  
 \hline
\end{tabular}

\end{table}
\end{center}



As shown above, the output sequence can be computed as $1111000$. This word output corresponds to the  polynomial $c(x) = 1 + x + x^2 + x^3$. 

This represents how a convolutional code can be encoded by constructing a shift register to obtain the output sequence graphically. 

We can also encode by computing the output polynomial $c(x) = a(x)g(x)$ directly. 

\begin{table}[!htp]\centering

\begin{tabular}{r l} 

$a(x)g(x)$ & = $(1 + x)(1 + x^2)$ \\

& $= 1 + x^2 + x + x^3$ \\

& $= 1 + x + x^2 + x^3$ \\

& $= c(x)$ \\

\end{tabular}

\end{table}

\end{exmp}


A codeword $c(x)=(c_1(x), c_2(x), \dots,c_n(x))$ can also be obtained in an interleaved form  such that the components $c_1(x), c_2(x), \dots,c_n(x)$ of the output  are combined where the $n$ digits that are  coefficients of the same term $x^i$ are grouped together. An additional way to encode convolutional codes is by utilizing Trellis Diagrams \cite{ViraktamathTrellis}. 

\begin{exmp}
The message $m(x) = 1 + x^3$ can be encoded using the $(3, 1, 3)$ convolutional code with generators $g_1(x) = 1 + x + x^3, g_2(x) = 1 + x + x^2 + x^3,$ and $g_3(x) = 1 + x^2 + x^3$. 

The shift register for this code is depicted below. 



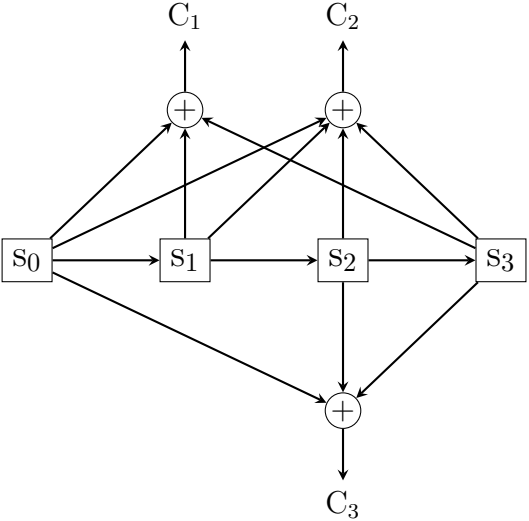
\begin{figure}[h!]
\begin{center}
\begin{tikzpicture}[node distance=2cm]
\node (s0)[rect01]{\large s$_0$};

\node (s1)[rect01,right of=s0, xshift=0.1cm]{\large s$_1$};
\node (s2)[rect01,right of=s1, xshift=0.1cm]{\large s$_2$};
\node (s3)[rect01,right of=s2, xshift=0.1cm]{\large s$_3$};

\draw [arrow] (s0) -- (s1);
\draw [arrow] (s1) -- (s2);
\draw [arrow] (s2) -- (s3);

\node (p1) [circ,above of= s1]{+};
\node (p2) [circ,above of= s2]{+};
\node (p3) [circ,below of= s2 ]{+};
\draw[arrow] (s0)--(p1);
\draw[arrow] (s1)--(p1);
\draw[arrow] (s3)--(p1);
\draw[arrow] (s0)--(p2);
\draw[arrow] (s1)--(p2);
\draw[arrow] (s2)--(p2);
\draw[arrow] (s3)--(p2);

\draw[arrow] (s0)--(p3);
\draw[arrow] (s2)--(p3);
\draw[arrow] (s3)--(p3);

\node (c1) [above of= p1, node distance=1.25cm]{C$_1$};
\node (c2) [above of= p2, node distance=1.25cm]{C$_2$};
\node (c3) [below of= p3, node distance=1.25cm]{C$_3$};

\draw [thick,-stealth](p1) -- (c1);
\draw [thick,-stealth](p2) -- (c2);
\draw [thick,-stealth](p3) -- (c3);


\end{tikzpicture}
\caption{Shift register for the (3, 1, 3) convolutional code with 3 generators. }
\end{center}
\end{figure}

As we have a $(3, 1, 3)$ convolutional code, this means it has $n = 3$ generators. Based on the Definition 3.1, the message polynomial $m(x) = 1 + x^3$ can be encoded as the following:

\begin{center}
\begin{table}[!htp]\centering

\begin{tabular}{r l} 

$c(x)$ &= $((1 + x^3)g_1(x), (1 + x^3)g_2(x), (1 + x^3)g_3(x))$ \\

& $ = ((1 + x^3)(1 + x + x^3), (1 + x^3)(1 + x + x^2 + x^3), (1 + x^3)(1 + x + x^2 + x^3))$ \\ 
& $ = (1 + x + x^4 + x^6, 1 + x + x^2 + x^4 + x^5 + x^6, 1 + x^2 + x^5 + x^6)$ \\
& $\longleftrightarrow (1100101..., 1110111..., 1010011...)$
\end{tabular}

\end{table}
\end{center}
\end{exmp}

\vspace{-1cm}

When utilizing convolutional codes with $k \geq 1$, there are two ways to interpret the shift registers: 1) $k$ digits of the message sequence are inputted into the shift register for each computed output, or 2) the shift register is divided into $k$ different streams, where each stream is inputted as an entire unit into the shift register. 

\begin{exmp}
Assume we want to encode the messages $m(x) = 1 + x + x^3 + x^4 + x^5$ using the (3, 2, 4) convolutional code with generators $g_1(x) = 1 + x^3, g_2(x) = x + x^4, $ and $g_3(x) = 1 + x + x^2 + x^3 + x^4$. As described above, there are two techniques of encoding we can use. 

The message polynomial $m(x) = 1 + x + x^3 + x^4 + x^5$ corresponds to the message sequence $m = 110111...$

The first interpretation of $k = 2$ is to encode $m$ using the single shift register shown below and to move $k = 2$ message digits into the shift register at a time. 


The contents of the registers and the outputs are summarized in the following table. 
\newpage
\begin{center}
\begin{table}\centering
\caption{Constructed register and output table of shifts. As $k = 2$, 2 input symbols are inputted at a time.}
 \begin{tabular}{|c c c c c c c c|} 
 \hline
  &  &  &  & &  Output  & &\\
 Time & Input & $X_0$ \space $X_1$ \space $X_2$ \space $X_3$ \space $X_4$ & & $c_1$  &  $c_2$  &  $c_3$  &\\
 \hline
 -1 & --- & 0 \space 0 \space 0 \space 0 \space 0   & &    &  &   &\\
 
 0 & 11 & 1 \space 1 \space 0 \space 0 \space 0   & &   1  & 1  & 0  &\\
 
 1 & 01 & 0 \space 1 \space 1 \space 1 \space 0   & &   1  & 1  & 1  &\\

2 & 11 & 1 \space 1 \space 0 \space 1 \space 1   & &   0 & 0  & 0  &\\

3 & 00 & 0 \space 0 \space 1 \space 1 \space 0  & &   1   & 0  & 0  &\\

4 & 00 & 0 \space 0 \space 0 \space 0 \space 1  & &   0   & 1  & 1 &\\

5 & 00 & 0 \space 0 \space 0 \space 0 \space 0  & &   0   & 0  & 0   &\\   [1ex]
 \hline
\end{tabular}

\end{table}
\end{center}

So, $m$ is encoded into the codeword (in interleaved form) 

\begin{center}
 $110 $ 111 $ 000 $ $100 $ $ 011$ $000$ ... 
\end{center}

The second interpretation of $k = 2$ is to notice that the first, third, fifth, ... message digits that feed into the shift register only ever appear in $X_0$, $X_2$, and $X_4$, and the second, fourth, sixth, ... message digits that feed into the shift register only appear in $X_1$ and $X_3$. Thus, the message and the registers can be split into $k = 2$ parts, as shown in Figure 4 below. 
\begin{figure}[h!]
\begin{center}
\begin{tikzpicture}[node distance=2cm, scale=0.5]

\node (m) {000000110111};

\node (x0)[rect01,right of=m,xshift=0.1cm]{\small X$_0$};

\node (x1)[rect01,right of=x0, xshift=0.1cm]{\small X$_1$};
\node (x2)[rect01,right of=x1, xshift=0.1cm]{\small X$_2$};
\node (x3)[rect01,right of=x2, xshift=0.1cm]{\small X$_3$};
\node (x4)[rect01,right of=x3, xshift=0.1cm]{\small X$_4$};

\draw [arrow] (m) -- (x0);
\draw [arrow] (x0) -- (x1);
\draw [arrow] (x1) -- (x2);
\draw [arrow] (x2) -- (x3);
\draw [arrow] (x3) -- (x4);

\node (p1) [circ,above of= x1, yshift=0.5cm, xshift = 1cm]{+};
\node (p2) [circ,above of= x3, yshift=0.5cm]{+};
\node (p3) [circ,below of= x2, yshift=-0.5cm]{+};


\draw[arrow] (x0)--(p1);
\draw[arrow] (x3)--(p1);

\draw[arrow] (x1)--(p2);
\draw[arrow] (x4)--(p2);

\draw[arrow] (x0)--(p3);
\draw[arrow] (x1)--(p3);
\draw[arrow] (x2)--(p3);
\draw[arrow] (x3)--(p3);
\draw[arrow] (x4)--(p3);

\node (c1) [above of= p1, node distance=1.25cm]{C$_1$};
\node (c2) [above of= p2, node distance=1.25cm]{C$_2$};
\node (c3) [below of= p3, node distance=1.25cm]{C$_3$};

\draw [thick,-stealth](p1) -- (c1);
\draw [thick,-stealth](p2) -- (c2);
\draw [thick,-stealth](p3) -- (c3);

\end{tikzpicture}
\caption{Shift register for the (3, 2, 4) convolutional code with three  generators.}
\end{center}
\end{figure}
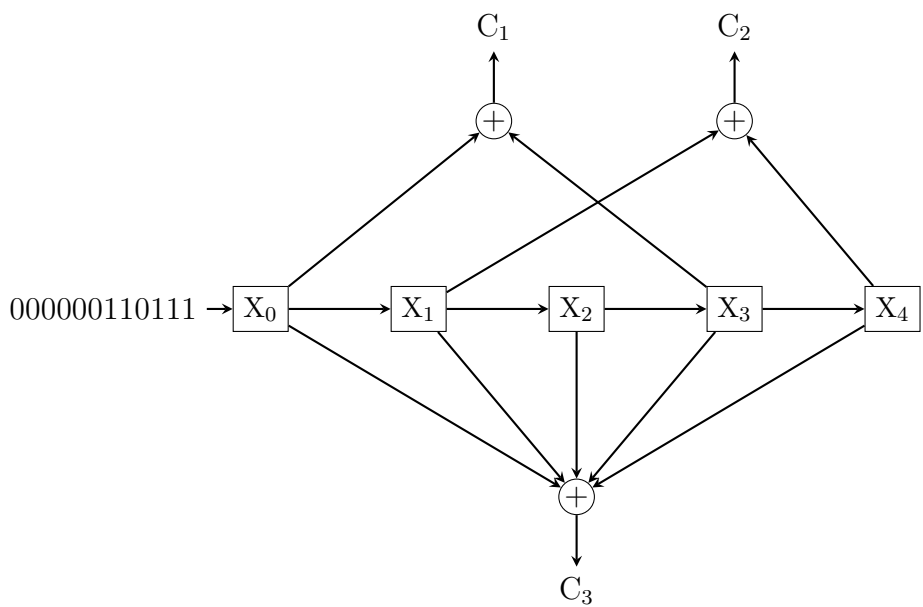

\pagebreak

\begin{figure}[h!]
\begin{center}
\begin{tikzpicture}[node distance=2cm]

\node (m1) {000111};

\node (x0)[rect01,right of=m1,xshift=0.1cm]{\small X$_0$};
\node (x2)[rect01,right of=x0, xshift=0.1cm]{\small X$_2$};
\node (x4)[rect01,right of=x2, xshift=0.1cm]{\small X$_4$};

\node (x1)[rect01,below of=x0,yshift=-1cm]{\small X$_1$};
\node (x3)[rect01,right of=x1, xshift=1cm]{\small X$_3$};

\node (m2) [left of=x1] {000101};

\draw [arrow] (m1) -- (x0);
\draw [arrow] (x0) -- (x2);
\draw [arrow] (x2) -- (x4);

\draw [arrow] (x1) -- (x3);

\draw [arrow] (m2) -- (x1);

\node (p1) [circ,above of= x4, yshift=0cm, xshift = 3cm]{+};
\node (p2) [circ,right of= x3, yshift=-1.5cm, xshift = 2cm]{+};
\node (p3) [circ,right of= x3, yshift=1.5cm, xshift = 2cm]{+};


\draw[arrow,green] (x0)--(p1);
\draw[arrow,green] (x3)--(p1);

\draw[arrow,red] (x1)--(p2);
\draw[arrow,red] (x4)--(p2);

\draw[arrow,blue] (x0)--(p3);
\draw[arrow,blue] (x1)--(p3);
\draw[arrow,blue] (x2)--(p3);
\draw[arrow,blue] (x3)--(p3);
\draw[arrow,blue] (x4)--(p3);

\node (c1) [right of= p1, node distance=1.25cm]{C$_1$};
\node (c2) [right of= p2, node distance=1.25cm]{C$_2$};
\node (c3) [right of= p3, node distance=1.25cm]{C$_3$};

\draw [thick,-stealth](p1) -- (c1);
\draw [thick,-stealth](p2) -- (c2);
\draw [thick,-stealth](p3) -- (c3);

\end{tikzpicture}
\caption{Adjusted shift register depicting the second interpretation of $k = 2$ where the first, third, fifth, etc. message digits that are fed into the shift register only appear in $X_0, X_2,$ and $X_4$ while the second, fourth, sixth, etc. message digits only appear in $X_1$ and $X_3$}
\end{center}
\end{figure}
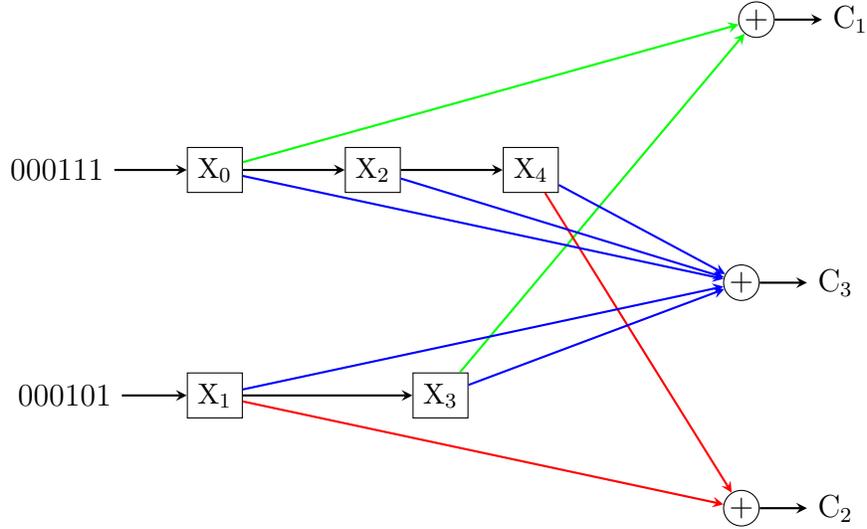

\end{exmp}

Another way to view the encoding of convolutional codes is through state diagrams and their respective tabular representations \cite{Hankerson}.

\begin{definition}
For $(n, k, m)$ convolutional codes, a state diagram can be defined as long as $1 \leq k \leq \frac{m}{2}$. Each state $s = s_k, s_{k+1}, ... , s_m$ contains binary words of length $m + 1 - k$. Each binary word $u$ of length $k$ has an edge directed from state $s$ to state $u, s_k, s_{k + 1}, ..., s_m$. For $k > 1$, assume $k$ message digits are shifted into a single shift register for each $n$ output bits. 
\end{definition}

For the special case where $k = 1$, a $(2, 1, m)$ convolutional code is encoded with a shift register that has $m + 1$ registers. The state of the shift register is the contents of the first $m$ registers at each time. The zero state is when each of first $m$ registers contains the value 0. Assume the shift register is currently in state $s_0, s_1, ... , s_{m-1}$. At the next time increment, the shift register can be in two possible states. If the next message digit is a 0, the shift register is in state $0, s_0, s_1, ... , s_{m-2}$. If the next message digit is a 1, the shift register is in state $1, s_0, s_1, ... , s_{m-2}$ \cite{Hankerson}. 

The state diagram of a $(2, 1, m)$ convolutional code is a graph in which states of binary codes of length $m$ are represented as vertices. 

For each state $s = s_1, s_2 ... , s_{m}$, there are two edges directed outwards \cite{Hankerson}. 

1. One edge is directed from $s$ to state $0, s_1, ... , s_{m}$, which is labeled with the output when the registers $X_0, X_1, ... , X_m$ contain $0, s_1, ... , s_{m}$

2. One edge is directed from $s$ to state $1, s_1, ... , s_{m}$ which is labeled with the output when the registers $X_0, X_1, ... , X_m$ contain $1, s_1, ... , s_{m}$

The tabular representation of the state diagram shows the current state, which is the contents of registers $X_0, X_1, ... , X_{m-1}$, and the corresponding outputs which depend on whether the next message digit $X_m$ is a 0 or a 1. 

As message digits are inputted one-by-one (since $k = 1$), into the shift register, they move to different states and the generator polynomials output the  digits in the codeword. In the state diagram, this corresponds with proceeding along the directed arrows to adjacent vertices or states, where the directed arrows are labeled with the outputs of the generator polynomials. Hankerson et al. (\cite{Hankerson})  refers to the outputted codeword as a $(directed)$ $walk$ in the state diagram, starting at the zero state, $000...$, and moving along arrows to adjacent states. Recovering the message that corresponds to codewords is simple as in the state diagram, each message digit inputted into the shift register is the first digit in the state to which the state register moves. 

\begin{exmp}
Let $C_1$ be the $(2, 1, 2)$ convolutional code with generators $g_1(x) = 1 + x^2$ and $g_2(x) = 1 + x + x^2$. The state diagram for this code can be described as follows.

\begin{figure}[h!]
\begin{center}
    
\begin{tikzpicture}[->,>=stealth',shorten >=1pt,auto,node distance=2.8cm,  semithick]
  \tikzstyle{every state}=[fill=white,draw=black,text=black]

  \node[state] (A)                    {$00$};
  \node[state]         (B) [above right of=A] {$10$};
  \node[state]         (D) [below right of=A] {$01$};
  \node[state]         (C) [below right of=B] {$11$};
 
  \path (A) edge [loop left] node {00} (A) 
            edge              node {11} (B)
        (B) edge [bend right] node {01} (D)
            edge              node {10} (C)
        (C) edge [loop right] node {01} (C) 
            edge              node {10} (D)
        (D) edge              node {11} (A)
            edge [bend right] node {00} (B);
\end{tikzpicture}
\caption{State diagram for (2, 1, 2) convolutional code with two generators. Nodes represent the different states, and the directed edges are labeled with the output sequences of each corresponding walk.}
\end{center}
\end{figure}
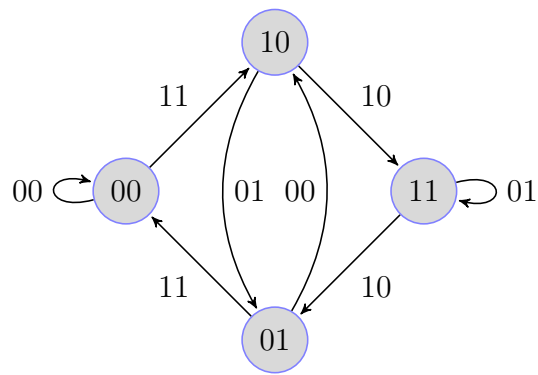

Each state $s = s_1, s_{2}$ in this diagram contains binary words of length $m = 2$. The four possible states, 00, 01, 10, 11, are represented as four vertices on the state diagram. The initial state of the shift diagram is the zero state 00. For the state $s = s_1s_2$ = 10, there is a directed edge from $s$ to the state $1s_1 = 11$. The directed edge from 10 to 11 is labeled with the output, based on the generator polynomial, where $X_0X_1X_2 = 110$, namely 10. 

Using the state diagram above, various messages can be encoded. 

Assume a message polynomial $m(x) = 1 + x^2$ which corresponds to the sequence 101....

Starting at the zero state, we move to the next state, 10, which corresponds to a 1 as the next input digit. The output value for this directed edge between the zero state 00 and 10 is 11, which is the initial part of the codeword. From 10, there are two possible paths to take: either going to 11 with 1 as the next input digit or traveling to 01 with 0 as the next input digit. The message sequence tells us that 0 is the next input digit, so the second directed segment is from 10 to 01 with the corresponding output sequence of 01. Continuing this directed walk, from 01, we travel back to 10 as that state contains the next message digit of 1. This directed walk corresponds to the output of 00. Completing this directed walk, we get the ultimate output codeword of 00, 11, 01, 00. Excluding the zero state, the corresponding encoded message is 11 01 00. 
\end{exmp}

The use of a state diagram for encoding can also be applied to a (2, 1, 3) convolutional code, as shown in the example below. 

\begin{exmp}
Let $C_1$ be the $(2, 1, 3)$ convolutional code with generators $g_1(x) = 1 + x + x^2 + x^3$ and $g_2(x) = 1 + x^2 + x^3$. The state diagram for this code can be created as follows. 

\begin{figure}[h!]
\begin{center}
    
\begin{tikzpicture}[->,>=stealth',shorten >=1pt,auto,node distance=3cm,  semithick, x=1cm, y=0.5cm]
  \tikzstyle{every state}=[fill=white,draw=black,text=black]


  \node[state] (A0)                     {$000$};
  \node[state] (A1) [below right of=A0, yshift = -\y, xshift=0cm] {$001$};
  \node[state] (A2) [right of=A0, xshift=1cm] {$010$};
  \node[state] (A3)  [right of=A1, xshift=3cm] {$011$};
  \node[state] (A4)  [above right of=A0, yshift = \y, xshift=0cm] {$100$};
  \node[state] (A5)  [right of=A0, xshift=3cm]{$101$};
  \node[state] (A6)  [right of=A4, xshift=3cm] {$110$};
  \node[state] (A7)  [below right of=A6, yshift = -\y, xshift=0cm] {$111$};
 
   \path (A0) edge [loop left]      node {00} (A0) 
              edge                  node {11} (A4)
         (A1) edge                  node {11} (A0)
              edge                  node {00} (A4)              
         (A2) edge  [bend right]    node [below]{00} (A5)
              edge                  node {11} (A1)
         (A3) edge                  node {11} (A5) 
              edge                  node {00} (A1)
         (A4) edge                  node {01} (A6)
              edge                  node {10} (A2)  
         (A5) edge   [bend right]   node [above]{01} (A2)
              edge                  node {10} (A6)   
         (A6) edge                  node {01} (A3)
              edge                  node {10} (A7) 
         (A7) edge [loop right]     node {01} (A7)
              edge                  node {10} (A3) ;

\end{tikzpicture}
\end{center}
\caption{State diagram for (2, 1, 3) convolutional code with two generators. Nodes represent the different states, and the directed edges are labeled with the output sequences of each corresponding walk.}
\end{figure}
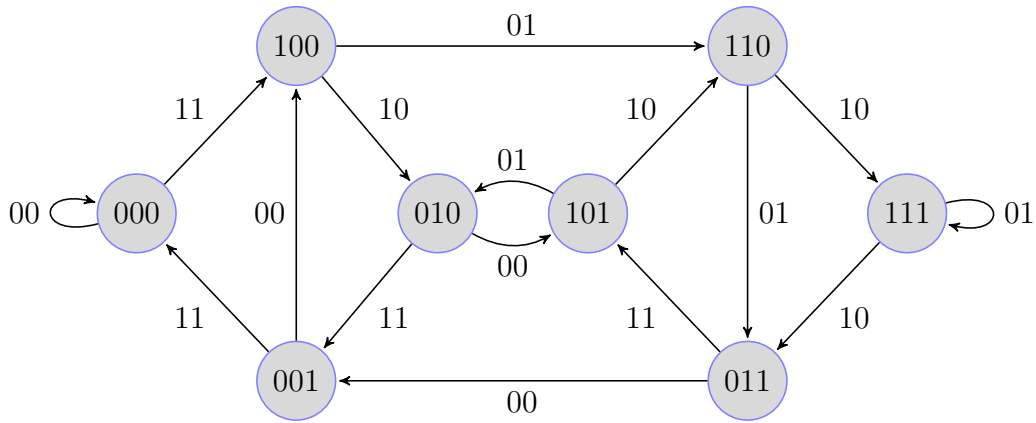

Each state $s = s_1, s_2, s_3$ contains binary words of length $m = 3$. Thus, the potential states, represented as vertices on the state diagram, are 000, 100, 010, 001, 101, 110, 011, 111. As the initial contents of the shift register contains 0, the initial state in the shift diagram is the zero state 000. The directed edges of the shift register are labeled with the outputs of each state transition based on the generator polynomial computations. For the state $s = s_1s_2s_3$ = 100, for example, there is a directed edge from $s$ to two states based on whether the next input digit is 0 or 1 respectively: $0s_1s_2$ and $1s_1s_2$. Both directed edges are labeled with the corresponding output sequence. When the next input digit is 0 and $X_0X_1X_2X_3 = 0100$, the output sequence is 10. When the next input sequence is 1 and $X_0X_1X_2X_3 = 1100$, the output sequence is 01. 

The state diagram can be used to encode and decode messages. 

Assume a message polynomial $m(x) = 1 + x + x^2 + ... = $$\sum_{i=0}^{\infty} x^{i} $$ $ which corresponds to the message sequence 111...

When encoding, the input message digits correspond to the first digit of the state to which the register moves to. Starting at the zero state 000, we move to the next state 100 based on the first message digit 1. This corresponds to the output value of 11, as written on the directed segment. Next, from 100, we move to 110 as the next message digit is 1. This corresponds to the output sequence of 01. Next, from 110, we move to 111, with an output sequence of 10. This process can be continued by taking 1s as inputs and continuing the walk in the state diagram. 

This directed walk corresponds to the codeword of 11 01 10 01 01 01 ... The state diagram can also be used to find the message corresponding to codewords, namely decoding. 

Let a partial codeword be 11 10 00 01 00 10 10... 

Starting at the zero state, recovering the message that corresponds to the codeword is fairly simple. Following the partial codeword as a directed codeword, we pass the following states: 000, 100, 010, 101, 010, 101, 110, 111. Taking the first digit of these states, we obtain the following message sequence that corresponds to the given codeword: 

1 0 1 0 1 1 1 ...
\end{exmp}

\subsection{Decoding Convolutional Codes}

As opposed to having a predetermined length, convolutional codewords have unbounded lengths, making decoding convolutional codes different from decoding other codes. Decoding begins before the entirety of the codeword is received in order to avoid storage problems. Determining the time to start decoding requires the introduction of a few concepts. 

In state diagrams, if a directed walk from the zero state to an output does not exactly match the received codeword, the decoded word that most likely fits the received codeword must be found. If there are storage restrictions or uncorrectable errors in transmission such that there are multiple disagreeing closest walks to the received codeword, the state is chosen arbitrarily and is denoted with an asterisk $*$. Conversely, if the entire correct codeword is given, it is fairly simple to find the closest codeword. 

Define $\tau$ as the window size or portion of the decoded word that is seen when making decoding decisions.

\begin{definition}
The exhaustive decoding algorithm for convolutional codes considers all walks of length $\tau$ from the current state to be considered for each message digit that is encoded. 
\end{definition}

So, if we wait $\tau$ steps before decoding, a decoding decision considers all walks of length $\tau$ from the current state, comparing each walk to the $\tau$ ticks of information from the current received word, and moving to the next state closest to the received word.

As shown, the amount of time waited before starting the decoding process affects the answer of the closest codeword. To determine the correct starting time, we must take a look at the error patterns that can be corrected. 

Define the weight of a walk or cycle through the state diagram as the (Hamming) weight of the outputs that are on the directed edges in the walk. Defined as the number of non-zero elements in a codeword, the Hamming weight is the most commonly used weight in coding theory. Whenever we use the term weight in this paper, we  mean Hamming weight 

\begin{definition}
A convolutional code is $catastrophic$ if its state diagram has a zero weight cycle different from the loop on the zero state. This paper assumes that the convolutional codes are not catastrophic. 
\end{definition}

Because of their linearity, convolutional codes can be characterized based on minimum free distance $d$, which is the weight of a non-zero codeword of least weight. As only non-catastrophic convolutional codes are taken into consideration, a non-zero codeword with finite weight corresponds to a walk that leaves the zero state, later returns to the zero state, and then stays there forever \cite{Hankerson}. For a non-catastrophic code, if a walk leaves the zero state, it must accumulate some positive weight as there are no zero weight cycles other than the zero state loop. 

\begin{exmp}
Take the code $C_1$ which is defined as a $(2, 1, 3)$ convolutional code with generator polynomials $g_1(x) = 1 + x + x^3$ and $g_2(x) = 1 + x^2+ x^3$. 

If there is a walk through the states 000 100 110 111 011 001 000 in the code's state diagram, the corresponding codeword is 11 01 00 00 10 11.  This corresponding codeword has a total weight of 6 as it contains  6 non-zero bits. All other walks in this code's state diagram also either have a weight equal to or greater than 6. This means that the minimum distance of the code is $d(C_1)$ = 6. 
\end{exmp}


If a code has a distance $d$, it should correct all error patterns of weight less than or equal to $\left [\frac{d - 1}{2} \right]$ \cite{Hankerson}.

The length of a walk is the number of directed edges in the walk that are enumerated each time they appear in the walk.

If all error patterns of weight less than or equal to $e$ are to be corrected, then the time $\tau(e)$ that must be waited before decoding should be long enough so that all walks of length $\tau(e)$ from the zero state that immediately leave the zero state have weight greater than 2$e$. 

So, suppose that the zero codeword is sent and less than or equal to $e$ errors occur during transmission. Due to the linearity of convolutional codes, there is no loss of generality in assuming that the zero codeword is sent. Using the exhaustive decoding algorithm with a window size of $\tau(e)$, after $\tau(e)$ ticks, the labels on all walks from the zero state that have length $\tau(e)$ to the first $\tau(e)$ ticks of the received word are compared, and then the closest walk is selected to determine which state should be moved to. To decode correctly, the decision should be to stay at the zero state as it is assumed that the zero codeword is sent. As a result of the choice of $\tau(e)$, all walks that immediately leave the zero state have weight greater than $2e$ after $\tau(e)$ steps, thus there is disagreement between the first $\tau(e)$ ticks of the received word $w$ in more than $e$ positions. The walk that never leaves the zero state has zero weight, so it has a distance $wt(w) \leq e$ from the first $\tau(e)$ ticks of $w$. None of the walks that immediately leave the zero state are the closest walks, so all closest walks to $w$ over the first $\tau(e)$ ticks agree that staying at the zero state is correct. No further decoding step is made until another tick of the codeword is received. When the new information is received, the same pattern can be repeated which shows that the received codeword $w$ can be decoded correctly. This actually proves that $w$ can be decoded correctly if less than or equal to $e$ errors occur in any $\tau(e)$ consecutive ticks of the received word. Infinitely many errors can be corrected as long as there are never more than $e$ errors in $\tau(e)$ consecutive ticks. This is similar to block codes that have finite lengths as errors are corrected provided less than or equal to $e$ errors occur in any codeword. 

This tells the time needed to wait before starting to decode.

\begin{definition}

A non-catastrophic convolutional code $C$, for $1 \leq e \leq \left [ \frac{d - 1}{2} \right]$, $
\tau(e)$ is defined as the smallest integer $x$ such that all walks of length $x$ in the state diagram that immediately leave the zero state have weight greater than $2e$. 
\end{definition}

The exhaustive decoding algorithm with window size $\tau(e)$ considers all walks of length $\tau(e)$, from the current state for each message digit to be be decoded. Constructing all $2^{\tau(e)}$ such walks at each time increment is time-consuming, so a faster algorithm is presented in \cite{Hankerson}. 

\begin{theorem} (\cite{Hankerson})
Let $C$ be a non-catastrophic convolutional code. For any $e$,\\ $1 \leq e \leq \left [{\frac{d - 1}{2}}\right ]$, if any error pattern containing less than or equal to $e$ errors in any $\tau(e)$ consecutive steps occurs during transmission, then the exhaustive decoding algorithm using the window size $\tau(e)$ will decode the received word correctly. 
\end{theorem}

This theorem states that if the exhaustive decoding algorithm is used with window size $\tau(1)$, all error patterns with less than or equal to $e = 1$ errors in any $\tau(1) = 2$ consecutive ticks will be corrected. For example, an error pattern such as $e_1 = $ 10 00 10 00 01 00 10 00 10 .. will be corrected. 



\subsection{Truncated Viterbi Decoding}

Hankerson presents a truncated Viterbi decoding algorithm for $(2, 1, m)$ binary convolutional codes. While the exhaustive decoding algorithm calculates and stores $2^\tau$ walks of length $\tau$ for each tick, the truncated Viterbi decoding algorithm calculates and stores $2^m$ walks of length $\tau$ at each time increment. The window size $\tau$ is set between $4m$ and $6m$ so that the value is larger than $\tau(e)$. Probabilistic arguments have shown that choosing a window size between these two values results in very few incorrectly decoded error patterns. Thus, storing $2^m$ walks instead of $2^\tau$ walks saves considerable time and space. 

The truncated Viterbi decoder is faster than the exhaustive decoding algorithm. In truncated Viterbi decoding, for each state $s$, less than or equal to one walk of length $\tau$ from the current state to $s$ is stored. On the other hand, in the exhaustive decoding algorithm, if the window size is $
\tau(e)$, then all walks of length $
\tau(e)$ need to be considered.

Let  $w = w_0, w_1, ..., w_i$ be the received word. For $i \geq 0, w_i$ is an $n$-tuple because codewords and received words are represented in interleaved form. Thus, in the case $n = 2$, $w_i$ has the 2 digits that are received at time $i$. 

For the first $m$ ticks, the decoder stores all walks from the zero state. When $t > m$ for each state $s = s_0, s_1, .., s_{m -1}$ there are two states that have directed edges to state $s$: 1) $S_0 = s_1, s_2, ..., s_{m - 1}, 0$ and 2) $S_1 = s_1, s_2, ..., s_{m -1}, 1$.At tick $t$, the distance between $w_t - 1$ and the outputs on directed edges from $S_0$ to $s$ and $S_1$ to $s$ are added respectively to $d(S_0, t - 1)$ and $d(S_1, t - 1)$. The smaller of the sums is $d(s; t)$ with the extension of the walk $W_0$ or $W_1$ that makes state $s$ the smaller distance. 

At time $m$, there are $2^m$ walks that each end in a different state. So this means that time $t = m$ is the first time there is exactly one walk that ends in each state. As the decoder builds the $2^m$ walks, it calculates how far the output of a walk is from the received word and stores the distance with the walk. At $t = m$, the decoder stores: 

1. The walk $W_0$ from the current state $s$ to $S_0$ as well as the distance $d(S_0:t)$ of $W_0$ to the received word. 

2. The walk $W_1$ from the current state $s$ to $S_1$ as well as the distance $d(S_1:t)$ of $W_1$ to the received word. 

Walks are stored as a sequence of message digits instead of a sequence of states or a sequence of outputs on the directed edges between states. Where $t \geq \tau$, a message digit is decoded at each time. The states that have the smallest distance function $d(s; t)$ are considered in two possible ways: 

1. If the walks that are stored in each state agree on which state to move to, meaning the walks have the same oldest message digit, then this message digit is decoded.

2. If the walks do not all agree, then the decoded message digit is flagged by decoding to $*$. It is also possible to arbitrarily decode to 0 in this situation, but it is helpful to see where neither the 0 or 1 message digit is the best. 

Once the message digit is decided, it can be removed from all stored walks. Thus, the length of the stored walks is reduced to $\tau - 1$, but it will increase back to $\tau$ when the walks are extended at the next time $t + 1$. 

\begin{Algorithm}\cite{Hankerson} Truncated Viterbi Decoding of a $(n, 1, m)$ convolutional code with a window size $\tau$ where $w_0w_1...$ is the received word. 

1. Initialization 

If $t = 0$, define 
\begin{center}
    $W(s; t) = s**...*$ of length $\tau$, and 

\end{center}
    \begin{equation}
  D_{it} =
    \begin{cases}
      0, & \text{if $s$ is the zero state, and}\\
      \infty, & \text{otherwise}\\
    \end{cases}       
\end{equation}

2. Distance Calculation 

For $t > 0$ and for each state $s = s_1, s_1, .., s_{m - 1},$  define distance 
\begin{center}
    $d(s; t)$  = min ${d(s_1, s_2, .., s_{m - 1}, 0; t - 1) + d_0(s),}$ \\ ${d(s_1, s_2, .., s_{m - 1}, 1; t - 1) + d_1(s)}$
\end{center}

where $d_i(s)$ is the distance between $W_{t-1}$ and the output on the directed edge \\
from $s_1, s_2, ... , s_{m - 1}$, $i$ to $s$. 

3. Walk Calculation 

a) If $d(s_1, ... , s_{m - 1}, i; t - 1) + d_i(s) < d(s_1, ... , s_{m - 1}, j; t - 1) + d_j(s), {i, j} = {0, 1}$, then form $W(s; t)$ from $W(s_1, ... , s_{m - 1}, i; t - 1)$ by adding the leftmost digit of $s$ to the left of $W(s_1, ... , s_{m - 1}, i; t - 1)$ and deleting the rightmost digit. 

b) If $d(s_1, ... , s_{m - 1}, 0; t - 1) + d_0(s) = d(s_1, ... , s_{m}, 1; t - 1) + d_1(s)$, form $W(s;t)$ from \\
$W(s_1, ... , s_m, 0; t- 1)$ by adding the leftmost digit of $s$ to the left of $W(s_1, ... , s_{m - 1}, 0; t - 1)$, replacing each digit that disagrees with $W(s_1, ... , s_{m - 1}, 1; t - 1)$ with $*$ and the deleting the rightmost digit. 

4. Decoding 

For $t \geq \tau$, let $S(t) = \{s| d(s; t) \leq d(s':t) \text{ for all states } s'\}.$ If the rightmost digit in $W(s;t)$ is the same value $i \text{ for all } s \in S(t)$, decode the message digit $i$. Otherwise, decode the message digit $*$.

The leftmost $m$ digits in $W(s; t)$ necessarily equal $s$, so they do not need to be stored. 

\end{Algorithm}

The last decoding step can also be defined in other ways. For example, it could be claimed that decoding should not take place until the walks to each state agree on the rightmost digit that is used for coding. But in this algorithm, we might need to wait for many ticks before any decoding can be done, which raises the issue of open ended storage requirements. 

Another possibility is to delete each walk in which the rightmost digit disagrees with the message digit currently being decoded as such walks choose to move to a different state. This decoding technique poses theoretical problems in the analysis of the algorithm, as it is conceivable that such a decoding algorithm might itself impose infinitely many decoding errors after a finite burst of errors during transmission. 















This algorithm can also be generalized to decode $(n, k, m)$ convolutional codes. 

\begin{Algorithm} \cite{Hankerson} Truncated Viterbi Decoding of a $(n, k, m)$ convolutional code with a window size $\tau$ where $w_0w_1...$ is the received word. 

1. Initialization 

If $t = 0$, define 
\begin{center}
    $W(s; t) = s**...*$ of length $\tau$, and 

\end{center}
    \begin{equation}
  D_{it} =
    \begin{cases}
      0, & \text{if $s$ is the zero state, and}\\
      \infty, & \text{otherwise}\\
    \end{cases}       
\end{equation}

This step is the same as in Algorithm 3.1. 

2. Distance Calculations 

For $t > 0$ and for each state $s = s_1, s_1, .., s_{m - k},$  define distance 
\begin{center}

 \bigskip
   $\displaystyle
 d(s; t)  = \min_{u} {d(s_k, .., s_{m - k}, u; t - 1) + d_u}$
\end{center}

where $u$ comprises all binary words of length $k$ and $d_u$ is the distance between $w_{t - 1}$ and the output on the directed edge from state $s_k, ... , s_{m - k}, u$ to state $s$ in the state diagram. 

3. Walk Calculations 

a) If $d(s_k, ... , s_{m - k}, u; t - 1) + d_u < d(s_k, ... , s_{m - k}, v; t - 1) + d_v, \forall v \neq u$, then form $W(s; t)$ from $W(s_k, ... , s_{m - k}, u; t - 1)$ by deleting the rightmost $k$ digits from it and adding the leftmost $k$ digits of $s$ to it. 

b) If $d(s_k, ... , s_{m - k}, u; t - 1)$ is not the smallest value for a unique choice of $u$, 
    
    i) Form $W(s; t)$ by choosing any such $u$ and proceeding in 3a. 
    
    ii) Take a combination of all the walks $W(s_k, ... , s_{m - k}, u; t - 1)$ for which $d(s_k, ... , s_{m - k}, u ; t - 1)$ is a minimum, placing an $*$ in any position where 2 such walks disagree. 
    
4. Decoding 

For $t \geq \tau$, let $S(t) = \{s| d(s;t) \leq d(s';t) \text{ for all states } s'\}$. If two walks agree, decode the leftmost digits $m_{1, t}, m_{2, t}, ... , m_{k, t}$ where $m_{i, t}$ is the $i$th digit in the rightmost $k$ digits of $W(s, t)$, for all $s \in S(t)$. If two walks disagree in the $i$th position, $m_{i, t} = *$
\end{Algorithm}

It is not possible for a result as strong as Theorem 3.1 to be proven for the truncated Viterbi decoder of Algorithm 3.1 as the truncated Viterbi decoding algorithm takes time to recover from errors in the codewords during transmission. Determining how long errors during transmission affect truncated Viterbi decoding is now explained.  

With window size $\tau(e)$, $w(s, s')$ is defined as the weight of the smallest weight path from $s$ to $s'$ in the state diagram. If $s(t)$ is the correct state at some tick $t$ (meaning $s(t)$ is the state the codeword sent is in at tick $t$), the decoder is defined to be $e$-ready at tick $t$ if the following conditions are met: 
 
1) $d(s';t) \geq d(s(t); t)$ + min ${ 1 + e, w(s(t), s')} \forall s' \neq s(t)$ 

2) if $w(s(t), s') < 1 + e$, then $W(s';t) = s'v$ of length $\tau$ where $v$ is defined by $W(s(t); t) = s(t)v$

\begin{theorem} (\cite{Hankerson})
If $C$ is a non-catastrophic convolutional code that is decoded through the truncated Viterbi decoder, at tick $t$, if the decoder is $e$-ready, then correct decoding will occur if less than or equal to $e$ errors are subsequently made during transmission. 
\end{theorem}

This result is not as strong as Theorem 3.1. A $guard space$ is the time period of error-free transmission following a burst of errors. To obtain a result that is comparable to Theorem 3.1, it is important to know how long a guard space is required before the decoder is $e$-ready. For the exhaustive decoder, the guard space required is 0, where $e-$ready means any subsequent error pattern of weight less than or equal to $e$ results in a correctly decoded received word. The guard space required for the truncated Viterbi decoder to become $e-$ready after a burst of errors is finite. The length of the guard space for a convolutional code with a small $m$ is known. The below theorem is the closest we can get to Theorem 3.1. 

\begin{theorem} (\cite{Hankerson})
Le $C$ be a non-catastrophic convolutional code which is decoded using the truncated Viterbi decoder with a window size $\tau(e)$. If the error pattern can be partitioned into bursts of errors, each of weight less than or equal to $e$ and each followed by a sufficiently long (finite) guard space, the decoder will decode correctly. 
\end{theorem}


The Viterbi decoding method was proposed by Andrew Viterbi in 1967 (\cite{viterbi}) and is still used in cell phones for error-correcting codes, speech recognition, DNA analysis, and applications of Hidden Markov models \cite{MIT Lecture 9}. An alternative way to encode and decode convolutional codes, in addition to the methods provided above, is through Trellis diagrams \cite{ViraktamathTrellis}. Assuming there are no errors, to decode using a Trellis diagram, one must follow a path through states of the diagram that match up with the received sequences \cite{MIT Lecture 8}. The corresponding output sequences will be labeled on the directed edges of the diagram and can be concatenated to create the output sequence. 

\begin{exmp}

Assuming a (2, 1, 2) convolutional code $C_1$ that has a memory order of $t = 2$. Let the received codeword be $w = w_0w_1w_2 = 01 11 10 10 00 11 10$. We can decode this received codeword by creating and following a Trellis diagram, as depicted below. 


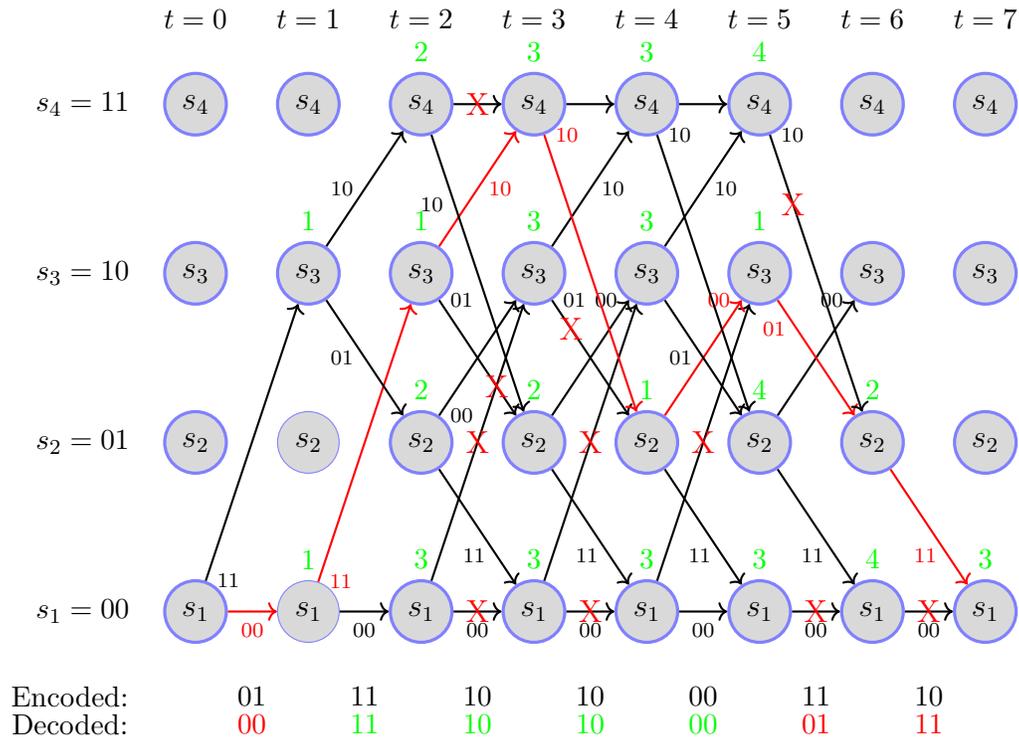
\begin{figure}[h!]

\begin{center}
\begin{tikzpicture}[scale=0.75]

\pgfmathsetmacro{\Z}{1}
\pgfmathsetmacro{\X1}{-6}
\pgfmathsetmacro{\X2}{-3}

\node               at (\z-8,\x+3.5*\y)[font=\small] {$t=0$};
\node               at (\z-10,\x+3*\y)[font=\small] {$s_4=11$};
\node               at (\z-10,\x+2*\y)[font=\small] {$s_3=10$};
\node               at (\z-10,\x+\y)[font=\small] {$s_2=01$};
\node               at (\z-10,\x) [font=\small] {$s_1=00$};

 
\node[mainstate] (s4_1) at (\z-8,\x+3*\y) {$s_4$} ;
\node[mainstate] (s3_1) at (\z-8,\x+2*\y) {$s_3$};
\node[mainstate] (s2_1) at (\z-8,\x+\y) {$s_2$};
\node[mainstate] (s1_1) at (\z-8,\x) {$s_1$};


\node               at (\z-6,\x+3.5*\y) [font=\small] {$t=1$};
\node[mainstate] (s4_2) at (\z-6,\x+3*\y) {$s_4$}    ;
\node[mainstate,label={[green,font=\small]above:1}] (s3_2) at (\z-6,\x+2*\y) {$s_3$}  edge[mainedge] node[at end, text=black,right,font=\scriptsize]{11} (s1_1);
\node[state] (s2_2) at (\z-6,\x+\y) {$s_2$}         ;
\node[state,label={[green,font=\small]above:1}] (s1_2)  at (\z-6,\x) {$s_1$}     edge[mainedge, red] node [below, font=\scriptsize]{00} (s1_1) ;

\node               at (\z-4,\x+3.5*\y) [font=\small] {$t=2$};

\node[mainstate,label={[green,font=\small]above:2}] (s4_3) at (\z-4,\x+3*\y) {$s_4$}   edge[mainedge] node[left,font=\scriptsize]{10}  (s3_2) ;
\node[mainstate,label={[green,font=\small]above:1}] (s3_3) at (\z-4,\x+2*\y) {$s_3$}  edge[mainedge, red] node[at end, right,font=\scriptsize]{11} (s1_2)    ;
\node[mainstate,label={[green,font=\small]above:2}] (s2_3) at (\z-4,\x+\y) {$s_2$}  edge[mainedge] node[left,font=\scriptsize]{01}(s3_2);    
\node[mainstate,label={[green,font=\small]above:3}] (s1_3) at (\z-4,\x) {$s_1$}  edge[mainedge] node [below,font=\scriptsize]{00} (s1_2)  ;

\node               at (\z-2,\x+3.5*\y) [font=\small] {$t=3$};

\node[mainstate,label={[green,font=\small]above:3}] (s4_4) at (\z-2,\x+3*\y) {$s_4$} edge[mainedge] node[text=red]{X} (s4_3) edge[mainedge, red] node[right,font=\scriptsize]{10} (s3_3);
\node[mainstate,label={[green,font=\small]above:3}] (s3_4) at (\z-2,\x+2*\y) {$s_3$}  edge[mainedge] node[text=red]{X} (s1_3) edge[mainedge] node [at end,right,font=\scriptsize]{00} (s2_3);
\node[mainstate,label={[green,font=\small]above:2}] (s2_4) at (\z-2,\x+\y) {$s_2$} edge[mainedge] node [at end,right,font=\scriptsize]{01} node[near start,text=red]{X} (s3_3) edge[mainedge] node[near end,left,font=\scriptsize]{10} (s4_3);
\node[mainstate,label={[green,font=\small]above:3}] (s1_4) at (\z-2,\x) {$s_1$} edge[mainedge] node [below,font=\scriptsize]{00} node[text=red]{X} (s1_3) edge[mainedge] node[near start, left,font=\scriptsize]{11} (s2_3);

\node               at (\z,\x+3.5*\y) [font=\small] {$t=4$};

\node[mainstate,label={[green,font=\small]above:3}] (s4_5) at (\z,\x+3*\y) {$s_4$}  edge[mainedge] (s4_4) edge[mainedge] node[right,font=\scriptsize]{10} (s3_4);
\node[mainstate,label={[green,font=\small]above:3}] (s3_5) at (\z,\x+2*\y) {$s_3$}  edge[mainedge] node[text=red]{X} (s1_4) edge[mainedge] node [at start,left,font=\scriptsize]{00} (s2_4);
\node[mainstate,label={[green,font=\small]above:1}] (s2_5) at (\z,\x+\y) {$s_2$} edge[mainedge] node [at end, right,font=\scriptsize]{01}  node[near end,text=red]{X} (s3_4) edge[mainedge, red] node[at end, right,font=\scriptsize]{10} (s4_4);
\node[mainstate,label={[green,font=\small]above:3}] (s1_5) at (\z,\x) {$s_1$} edge[mainedge] node [below,font=\scriptsize]{00} node[text=red]{X} (s1_4)  edge[mainedge] node[near start, left,font=\scriptsize]{11} (s2_4);

\node               at (\z+2,\x+3.5*\y)  [font=\small]{$t=5$};

\node[mainstate,label={[green,font=\small]above:4}] (s4_6) at (\z+2,\x+3*\y) {$s_4$} edge[mainedge] (s4_5) edge[mainedge] node[right,font=\scriptsize]{10} (s3_5);
\node[mainstate,label={[green,font=\small]above:1}] (s3_6) at (\z+2,\x+2*\y) {$s_3$}  edge[mainedge] node[text=red]{X} (s1_5) edge[mainedge, red] node [at start,left,font=\scriptsize]{00} (s2_5);
\node[mainstate,label={[green,font=\small]above:4}] (s2_6) at (\z+2,\x+\y) {$s_2$}  edge[mainedge] node [left,font=\scriptsize]{01}  (s3_5) edge[mainedge] node[at end,right,font=\scriptsize]{10} (s4_5);
\node[mainstate,label={[green,font=\small]above:3}] (s1_6) at (\z+2,\x) {$s_1$} edge[mainedge] node [below,font=\scriptsize]{00} (s1_5)  edge[mainedge] node[near start,left,font=\scriptsize]{11} (s2_5);

\node               at (\z+4,\x+3.5*\y)  [font=\small] {$t=6$};

\node[mainstate] (s4_7) at (\z+4,\x+3*\y) {$s_4$};
\node[mainstate] (s3_7) at (\z+4,\x+2*\y) {$s_3$} edge[mainedge] node [at start,left,font=\scriptsize]{00} (s2_6);
\node[mainstate,label={[green,font=\small]above:2}] (s2_7) at (\z+4,\x+\y) {$s_2$}  edge[mainedge, red] node [near end, left,font=\scriptsize]{01} (s3_6) edge[mainedge] node[at end,right,font=\scriptsize]{10} node[near end,text=red]{X} (s4_6);
\node[mainstate,label={[green,font=\small]above:4}] (s1_7) at (\z+4,\x) {$s_1$} edge[mainedge] node [below,font=\scriptsize]{00} node[text=red]{X} (s1_6) edge[mainedge] node[near start,left,font=\scriptsize,font=\scriptsize]{11} (s2_6);

\node               at (\z+6,\x+3.5*\y) [font=\small] {$t=7$};

\node[mainstate] (s4_8) at (\z+6,\x+3*\y) {$s_4$};
\node[mainstate] (s3_8) at (\z+6,\x+2*\y) {$s_3$};
\node[mainstate] (s2_8) at (\z+6,\x+\y) {$s_2$};
\node[mainstate,label={[ green,font=\small]above:3}] (s1_8) at (\z+6,\x) {$s_1$} edge[mainedge] node [below,font=\scriptsize]{00} node[text=red]{X} (s1_7)  edge[mainedge, red] node[near start,left,font=\scriptsize]{11} (s2_7);

\node at (-8,0) [left,font=\small] {Encoded:};

\node at (-6,0) [font=\small] {01};
\node at (-4,0) [font=\small]{11};
\node at (-2,0) [font=\small]{10};
\node at (0,0) [font=\small]{10};
\node at (2,0) [font=\small]{00};
\node at (4,0) [font=\small]{11};
\node at (6,0) [font=\small]{10};

\node at (-8,-0.5)  [left,font=\small]{Decoded:};

\node at (-6,-0.5) [red,font=\small]{00};
\node at (-4,-0.5)[green,font=\small] {11};
\node at (-2,-0.50)[green,font=\small] {10};
\node at (0,-0.50)[green,font=\small] {10};
\node at (2,-0.50)[green,font=\small] {00};
\node at (4,-0.50)[red,font=\small] {01};
\node at (6,-0.50)[red,font=\small] {11};

\end{tikzpicture}
\caption{Trellis diagram for the code $C_1$. The nodes depict the different possible states at varying times, the labels on the directed edges represent the outputs based on the generator polynomial, and the numbers in green depict the Hamming distances of the corresponding paths.}
\end{center}
\end{figure}

\end{exmp}




\section{DNA Codes}


Deoxyribonucleic acid, or DNA, is the biological material present in every living organism. DNA carries genetic information and codes for phenotypic features. A strand of DNA can be encoded with an arbitrary message and then chemically synthesized as oligonucleotides which can be transported, stored, or used \cite{Press}. Techniques such as DNA sequencing can be used to recover the digital message, ideally with minimal or no errors. Although molecular barriers are present to prevent extensive damage from errors in DNA sequences, DNA contains special cellular machinery designed for error-prevention and error-correction \cite{Battail}. Whether errors in DNA sequences arise from mutagenesis or molecular mishaps, the genome acts as an error-correcting code and is regenerated often to ensure the quantity of errors does not exceed error-correcting ability. 

A DNA strand is a helical polymer with a sequence of monomer units known as nucleotides. Each nucleotide consists of a phosphate group, a deoxyribose sugar, and a nitrogenous base. There are four nitrogenous bases in DNA: Cytosine (C), Guanine (G), Adenine (A), and Thymine (T). Each nucleotide is organized sequentially and forms a strand of DNA with chemically distinct  $5'$ and $3'$ ends and the double helix conformation of DNA is formed through specific hybridization \cite{Marathe}. The Watson-Crick canonical base-pairing of DNA states that Adenine (A) pairs with Thymine (T), while Guanine (G) pairs with Cytosine (C) \cite{Cleaves}. Biochemist Erwin Chargaff established the base equivalent rule stating the quantity of A is equivalent to T and the quantity of G is equivalent to C in human DNA strands \cite{Chargaff}.

\begin{exmp}
If we are told that A comprises $12 \% $ of the DNA strand, we can compute the breakdown of the DNA strands. From Watson-Crick base-pairing rules and Chargaff's rules, we know that the amount of T is equivalent to the amount of A. Thus, T also comprises $12 \% $ of the DNA strand. This leaves $76 \%$ of the DNA strand which is represented equally by C and G. In total, A comprises $12 \% $, T comprises $12 \% $, C comprises $38 \% $, and G comprises $38 \% $  of the DNA strand. 
\end{exmp}

\begin{exmp}
Based on the base-pairing rules of DNA, the complement of $5' - CATAGCGAT - 3'$ is $3' - GTATCGCTA - 5'$. 
\end{exmp}

DNA codewords are short DNA strands that store retrievable information \cite{Marathe}. As there are four nitrogenous bases in DNA, DNA codes use a four-letter alphabet, namely A, C, T, G, where a strand of length $n$ can represent up to $4^n$ values \cite{Limbachiya} \cite{L. P. Dinu} \cite{Marathe}.


DNA codes comprise a set of DNA codewords of length $n$ over the DNA alphabet $ $$\sum_{A, C, T, G} = {A, C, T, G} $$ $ with a predefined distance $d$. DNA codewords also aim to satisfy a set of constraints \cite{Limbachiya}. Our treatment of DNA codes is largely based on the definitions set forth by Limbachiya et. al in  ``The Art of DNA Strings: Sixteen Years of DNA Coding Theory'' \cite{Limbachiya}.

\begin{definition}
A DNA code of length $n$, size $M$ and minimum distance $d$ is\\ \calligra C\normalfont $_{DNA}(n, M, d)\subseteq  \sum^n$ where $\sum = \{A, C, T, G\} $. 
\end{definition}

\subsection{Constraints for the Construction of DNA Codes}

The three main types of constraints that DNA codes need to satisfy include combinatorial, thermodynamic, and application oriented constraints \cite{Limbachiya}. These constraints are summarized from \cite{Limbachiya}.


\begin{table} [h]
\caption{The combinatorial, thermodynamic, and application oriented constraints of DNA codes \cite{Limbachiya}.
}
\begin{center}
    \begin{tabular}[t]{|l|l|l|} \hline
    \multicolumn{3}{|c|}{\textbf{\scriptsize Constraints on DNA Codes}} \\ \hline
   \textbf{\scriptsize Combinatorial} & \textbf{\scriptsize Thermodynamic} & \textbf{\scriptsize Application-Oriented }\\ \hline
      {\scriptsize Hamming Distance (HD) Constraint} & {\scriptsize Melting Temperature Constraint} & {\scriptsize Run Length Constraint} \\ \hline
     {\scriptsize Reverse Constraint} & {\scriptsize Energy Minimization Constraint} & {\scriptsize Correlated-Uncorrelated Constraint} \\ \hline
     {\scriptsize Reverse Complement (RC) Constraint} & {\scriptsize Free Energy Constraint} &\\ \hline
     {\scriptsize GC-Content Constraint} & &  \\ \hline
     {\scriptsize Forbidden Constraint} & & \\ \hline
    \end{tabular}
\end{center}

\end{table}

Combinatorial constraints are based on Hamming Distance (HD) constraint, Reverse constraint, Reverse Complement (RC) constraint, GC-content constraint, and the Forbidden constraint. The HD, Reverse, and RC constraint are used to avoid undesirable hybridization, or joining, of differing DNA strands. 

Thermodynamic constraints include melting temperature, energy minimization, and free energy constraints. 

Application oriented constraints include the run length constraint and the correlated-uncorrelated constraint. 

Below is an overview of the above constraints supplemented with examples. 

\textbf{1. The Hamming Distance Constraint}

Given  $d$,  pairs of distinct DNA words $x, y \in $\calligra C\normalfont$_{DNA}$ satisfy the  Hamming distance constraint if
$d_H(x, y) \geq d$ $\forall$ $x, y \in $\calligra C\normalfont$_{DNA}$. A set of codewords of length $n$, size $M$, and minimum Hamming distance $d$, satisfying the Hamming constraint is represented by \calligra C\normalfont$_{DNA}(n, M, d)$. For  \calligra C\normalfont$_{DNA}(n, M, d)$, the Hamming distance is the least of the distances between  two DNA codewords. $A_q(n, d)$ is defined as the maximum size of a code with words of length $n$ over the alphabet size $q$. For DNA codes, $q = 4$, and $A_4(n, d)$ is the maximum size. 

\begin{exmp}
If $x = CTAGCT$ and $y = CATCGT$, then $d_H(x, y)$ $= 4$ where $x, y \in $\calligra C\normalfont$_{DNA}$, following the Hamming distance constraint with $d_H(x, y) \geq 3$.
\end{exmp}

\textbf{2. The Reverse Constraint}

For pairs of words $x, y \in $\calligra C\normalfont$_{DNA}$, the reverse constraint is $H_{DNA}(x^R, y)$ $\geq d$,  $\forall$ $x, y \in \calligra C\normalfont_{DNA}$, where $x^R$ is the reverse sequence of $x$ Reverse codes are codes that satisfy this reverse constraint. 

$A_q^R(n, d)$ is defined as the maximum size of a reverse code with words of length $n$ and a minimum Hamming distance of $d$.

\begin{exmp}
For $n = 6$ and $d = 3$, take $x = ACGATA$, $x^R = ATAGCA,$ and $y = TCTGGA$. The reverse constraint $H_{DNA}(x^R, y) = 4 \geq 3$ is satisfied, so $x$ and $y$ are in the reverse code. 
\end{exmp}

\textbf{3. The Reverse Complement (RC) Constraint}

 Given a distance parameter $d$, the RC constraint is satisfied by $C$ if $H(x^R,  y^C) \geq d$ for all $x, y \in $\calligra C\normalfont$_{DNA}$  where $x$ may equal $ y$.
 A reverse complement (RC) code is a DNA code that satisfies the RC-constraint. In other words, the reverse-complement of a DNA strand is constructed by first reversing the order of the bases and then substituting the bases with their complements \cite{Bishop}.

$A_q^{RC}(n, d)$ is defined as the maximum size of a reverse code with words of length $n$ and a minimum Hamming distance of $d$.

\begin{exmp}
For $n = 6$ and $d = 3$, take $x = ACGATA$, $x^R = ATAGCA$ and $y = TCTGGA, y^C = AGGTCT$. The reverse complement constraint $H_{DNA}(x^R,  y^C) = 4 \geq 3$ is satisfied, so $x$ and $y$ are in the reverse complement code. 
\end{exmp}

\begin{exmp} 
If a given DNA code has the sequence $ATAAGCT$, its reverse complement can be calculated as follows. 

First, by reversing the order of the bases, we get $TCGAATA$. 

Then, by substituting the bases with their complements, we arrive at the reverse complement sequence $AGCTTAT$.
\end{exmp}

\textbf{4. GC-Content Constraint}

The GC weight $w$ of a DNA strand is the total number of $G$s and $C$s present in the strand, namely $w_{x} = \#  \{x_i : x = x_i, x_i \in \{G, C\}  \}$ \cite{Chee} \cite{Tulpan} \cite{Smith}.  Generally, $w = \left [{\frac{n}{2}}  \right] $ \cite{Limbachiya}.

This constraint, alongside the thermodynamic constraint (6), ensures that all codewords  have relatively similar thermodynamic attributes, thus allowing for uniform computation.

\begin{exmp}
For $n = 6$, $x = ATAGGC \in $\calligra C\normalfont$_{DNA}$ because it satisfies the GC-content constraint $w = \left [ {\frac{n}{2}}\right ] = 3$. On the other hand, $x = ATAGGT \notin $\calligra C\normalfont$_{DNA}$ because it does not satisfy the GC-content constraint $w = \left [ {\frac{n}{2}}\right]  = 3$. 
\end{exmp}

\textbf{5. The Forbidden Constraint}

A DNA strand should not contain any undesired subsequences along the whole strand, at the $3'$ or $5'$ strands, or in the middle of the strand \cite{Tulpan-Design}. 

\textbf{6. Melting Temperature Constraint}

The melting temperature $T_m$ is the temperature at which half the DNA strands are hybridized and half are not \cite{Limbachiya}. Hybridization, the combination of strands, is the opposite of melting, in which the two strands are separated. Having codewords with similar melting temperatures is beneficial as it allows for the simultaneous hybridization of multiple DNA strands. Codewords with similar melting temperatures are selected based on Nussinov's algorithm \cite{Milenkovic}.

\textbf{7. Thermodynamic Constraint}
For DNA stability, all DNA codes should have comparable free energy $\Delta G\degree$ above a certain threshold \cite{Limbachiya}. DNA with minimum free energy $\Delta G\degree$ is relatively more stable, therefore only DNA codewords in the DNA code that have comparable free energy are considered. For DNA codewords $x, y \in$\calligra C\normalfont$_{DNA}$ and given a constant $\delta > 0$, the free energy $\Delta G\degree$ can be attained by the equation $\abs{\Delta G\degree(x) - \Delta G\degree(y)} \leq \delta$.

This constraint, alongside the GC-content constraint (4), ensures that all codewords  have relatively similar thermodynamic attributes for uniform computation. 

\textbf{8. Uncorrelated-Correlated Constraint}

This constraint states that a DNA codeword $(n, d, w) \in $\calligra C\normalfont$_{DNA}$ shifted by $x$ units where $x \leq n$ should not match any other codewords $\in $\calligra C\normalfont$_{DNA}$ \cite{Yazdi}.

We can define $x \circ y$ as the output of whether the $x$ and $y$ base pairs match up after each cyclical shift. When $y$ matches with $x,$ the output is 1 while if $y$ does not match with $x,$ the output is 0. This is what is represented by the rightmost column in the diagram below.  

\begin{exmp}
For $n = 7$, if $x = CGCATAC$ and $y = GCATACT$, then $x \circ y = 010000$, as shown below.

\begin{center}
 \begin{tabular}{c c c c c c c c c c c c c c c c c c c c} 
 X = & C & G & C & A & T & A & C & & & & & & & \\
 Y = & G & C & A & T & A & C & T & & & & & & 0 \\
 Y = & & G & C & A & T & A & C & T & & & & & 1 \\
 Y = & & & G & C & A & T & A & C & T & & & & 0 \\
 Y = & & & & G & C & A & T & A & C & T & & & 0\\
 Y = & & & & & G & C & A & T & A & C & T & & 0 \\
 Y = & & & & & & G & C & A & T & A & C & T & 0 \\
\end{tabular}
\end{center}
\end{exmp}

\subsection{Construction of DNA Codes}

Limbachiya et al. consolidate various algorithmic, theoretical, and software simulation approaches for constructing DNA codes with length $n$, distance $d$, and a set of constraints based on the specific application. Optimal construction of DNA codes is achieved when a maximum number of constraints is met for ``a large value of $n$ and large minimum distance $d$ with minimum error in DNA computation'' \cite{Limbachiya}. 


\begin{figure}[h]
\begin{tikzpicture}[scale=0.9, sibling distance=5em,
  every node/.style = {scale=.45, shape=rectangle, rounded corners,
    draw, align=center,
    top color=white, bottom color=white,text width=3cm,
}]]

  \node {DNA Codes Construction}
    child { node {Search Algorithms} 
        child { node {Variable Neighborhood Approach}
            child { node {Seed Building}}
            child { node {Clique Search}}
            child { node {Hybrid Search}}
            child { node {Greedy Approach}}
            child { node {Lexicographic Approach}}}
        child { node {Simulated Annealing}}
        child { node {Stochastic Local Search}}}
    child { node {Template Based Method}}
    child{ node {Algebraic Number Theory Codes}}
    child{ node {Algebraic Codes}
        child { node {Rings}}
        child { node {Linear Codes over GF(4)}}}
    child{ node {Altruistic Approach} } 
    child { node {Software Simulation}
        child { node {DNA Sequence Generator}}
        child { node { PUNCH (Princeton)}}};
\end{tikzpicture}
\caption{Tree diagram of the methods of DNA code construction }
\end{figure}
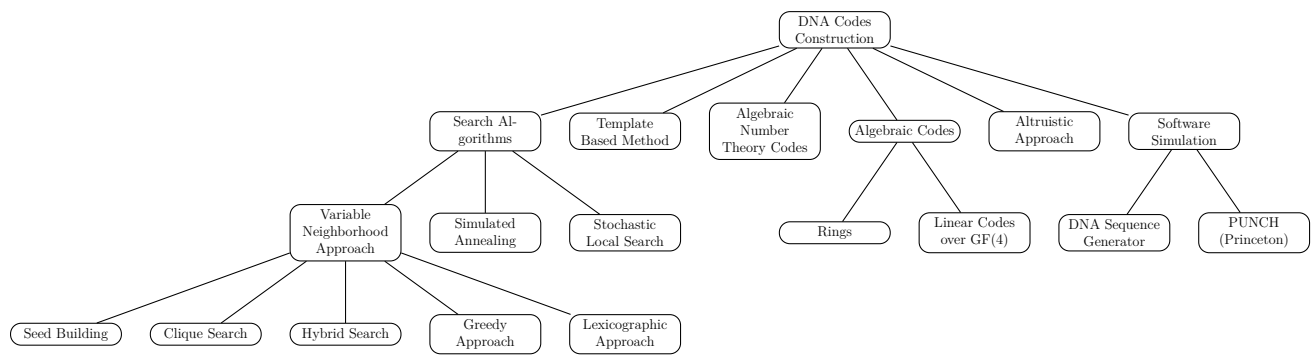

\subsection{Binary Representation of DNA Codes}
 








DNA codes are normally constructed using the four-letter alphabet A, C, T, G. As these DNA sequences go through the central dogma of genetics, they are encoded to mRNA sequences and then their corresponding amino acid sequences. Through this process, these sequences can form 64 possible triplet codons. In mRNA, Thymine (T) is replaced with Uracil (U), so the four possible nucleotide bases are A, C, U, and G. In a study published in 2017 ( \cite{Nemzer}), Nemzer devised a process for a binary representation of the genetic code, in which the four mRNA nucleotides, A, C, U, and G, were assigned unique two-bit identifiers, which were designated based on the molecular structures, physiological meanings, and relationships between these bases. One particular characteristic that sets Nemzer's method of creating a binary representation for DNA codes apart is that it is not arbitrary and is instead based on biological principles. This system takes into account the biological conjecture that genetic code is modified by natural selection, its evolution, the DNA mutation repair system, and the tRNA translation mechanism \cite{Nemzer}. Amino acids that are most ``determinative'' or critical for proper protein folding and function are placed first, following nature's laws \cite{Nemzer}. 

The methodology Nemzer provides is as follows. 

A set of four elements can be divided into $\binom{4}{2} = 6$ unique pairs. 

Digitally, we can represent these four states as 00, 01, 10, 11.  

For the 64 total codons, an amino acid correspondence table with 6-bit indices was developed \cite{Nemzer}. Each base is assigned to a 2-bit identifier, based on its properties. The first bit designates whether the base is a pyrimidine (0) or purine (1), and the second bit designates whether the base is weak (forms two bonds during Watson-Crick pairing) (0) or strong (forms three bonds during pairing) (1) \cite{Nemzer}. The following assignments are made. 

\begin{center}
U = 00 \\
C = 01 \\
A = 10\\
G = 11\\
\end{center}

Then, a truth table depicting the possible combinations of the six pairs is created. Each pair is given a binary representation based on whether it is a purine or pyrimidine, is weak or strong, or is a keto (U or G) or amino (A or C). The exclusive or (XOR) operation is applied such that 0 is outputted for keto bases as they have both bits as either 0 or 1, and 1 is outputted for amino bases as they have different bits.  

\begin{center}
\begin{table}[!htp]\centering
\caption{Truth table depicting combinations.}
 \begin{tabular}{|c | c c c c |} 
 \hline
 & U & C & A & G \\
 & 00 & 01 & 10 & 11 \\
 \hline
Pyrimidine (0) & 1 & 1 & 0 & 0\\
Purine (1) & 0 & 0 & 1 & 1\\
Weak (0) & 1 & 0 & 1 & 0 \\
Strong (1) & 0 & 1 & 0 & 1 \\
Keto (XOR = 0) & 1 & 0 & 0 & 1 \\
Amino (XOR = 1) & 0 & 1 & 1 & 0 \\

 \hline
\end{tabular}
\end{table}
\end{center}

The 64 tri-nucleotide codons, the units of mRNA that are translated to proteins, are organized into a table. The second bases in codons have been determined to convey the most information while the third base is often considered ``degenerate,'' due to the phenomenon known as wobble pairing \cite{Nemzer}. Thus, the order of importance of the bases in a codon is 2, 1, 3. 

Following this, the start codon, AUG, can be written in binary representation with a 6-bit index as follows. The 2-bit identifiers for the codons are concatenated in the 2, 1, 3, base order. So, for AUG, the order is U (00), A (10), and G (11), giving 001011. This method is followed to assign each of the 64 codons their respective 6-bit index which puts the most significant information bits first. The amino acid correspondence table is depicted below. 

\begin{table}[!htp]\centering
\caption{Amino Acid Binary Correspondence Table}\label{tab: }
\scriptsize
\begin{tabular}{lrrrrrrrrrrrrrr}\toprule
\multirow{2}{*}{\textbf{1st Base}} &\multicolumn{12}{c}{\textbf{2nd Base}} &\multirow{2}{*}{\textbf{3rd Base}} \\\midrule
&\multicolumn{3}{c}{\textbf{U}} &\multicolumn{3}{c}{\textbf{C}} &\multicolumn{3}{c}{\textbf{A}} &\multicolumn{3}{c}{\textbf{G}} & \\
\multirow{4}{*}{\textbf{U}} &UUU &000000 &Phe &UCU &010000 &Ser &UAU &100000 &Tyr &UGU &110000 &Cys &\textbf{U} \\
&UUC &000001 &Phe &UCC &010001 &Ser &UAC &100001 &Tyr &UGC &110001 &Cys &\textbf{C} \\
&UUA &000010 &Leu &UCA &010010 &Ser &UAA &100010 &Stp &UGA &110010 &Stp &\textbf{A} \\
&UUG &000011 &Leu &UCG &010011 &Ser &UAG &100011 &Stp &UGG &110011 &Trp &\textbf{G} \\
\multirow{4}{*}{\textbf{C}} &CUU &000100 &Leu &CCU &010100 &Pro &CAU &100100 &His &CGU &110100 &Arg &\textbf{U} \\
&CUC &000101 &Leu &CCC &010101 &Pro &CAC &100101 &His &CGC &110101 &Arg &\textbf{C} \\
&CUA &000110 &Leu &CCA &010110 &Pro &CAA &100110 &Gln &CGA &110110 &Arg &\textbf{A} \\
&CUG &000111 &Leu &CCG &010111 &Pro &CAG &100111 &Gln &CGG &110111 &Arg &\textbf{G} \\
\multirow{4}{*}{\textbf{A}} &AUU &001000 &Ile &ACU &011000 &Thr &AAU &101000 &Asn &AGU &111000 &Ser &\textbf{U} \\
&AUC &001001 &Ile &ACC &011001 &Thr &AAC &101001 &Asn &AGC &111001 &Ser &\textbf{C} \\
&AUA &001010 &Ile &ACA &011010 &Thr &AAA &101010 &Lys &AGA &111010 &Arg &\textbf{A} \\
&AUG &001011 &Met &ACG &011011 &Thr &AAG &101011 &Lys &AGG &111011 &Arg &\textbf{G} \\
\multirow{4}{*}{\textbf{G}} &GUU &001100 &Val &GCU &011100 &Ala &GAU &101100 &Asp &GGU &111100 & Gly &\textbf{U} \\
&GUC &001101 &Val &GCC &011101 &Ala &GAC &101101 &Asp &GGC &111101 &Gly &\textbf{C} \\
&GUA &001110 &Val &GCA &011110 &Ala &GAA &101110 &Glu &GGA &111110 &Gly &\textbf{A} \\
&GUG &001111 &Val &GCG &011111 &Ala &GAG &101111 &Glu &GGG &111111 &Gly &\textbf{G} \\
\bottomrule
\end{tabular}
\end{table}

\section{Applications of Convolutional Codes in DNA Computing}

\subsection{The Rise of Convolutional Codes in DNA Codes}

Error-correction in DNA codes has been a topic of great interest in the field of biocomputation. The search for effective coding theory models for DNA sequences that successfully take into consideration relevant biological phenomena is very timely. Various models based on communication theory have been developed to parallel DNA processes. We will now discuss previous models and studies.

Information storage and retrieval in the genetic code is also an important topic concerning biological coding theory. As short DNA strands are efficiently synthesized and can represent up to $4^n$ values, DNA codewords are optimal for molecular information storage and provide a basis for biocomputation \cite{Adleman} \cite{Marathe}.

Two early, pertinent studies in the fields of biocomputation are those done by May et al (\cite{May}) and Ponnala et al (\cite{Ponnala}) in the early 2000s in which the initiation of mRNA translation in E. coli K-12 was analyzed through block code models as well as rudimentary convolutional code models. While these studies had better results using the block code models, they demonstrated the potential of convolutional code models that should be studied further \cite{Liu}. These studies are explored further in Section 5.3 of this paper.  



\subsection{Why Convolutional Codes are Relevant to DNA Codes}


There are various reasons why convolutional codes, as opposed to block codes, are especially relevant to biological coding theory. Convolutional codes optimally take into account relevant biological phenomena as pertaining to DNA sequences and the central dogma of genetics. 

1. The phenomenon of codon context describes how adjacent nucleotides, containing $A, T, C, G$, in a DNA sequence affect the expression and efficiency of codon translation \cite{Shpaer} \cite{Yarus}. A codon has been shown to carry not only its own genetic information, but also parts of the genetic and error-correcting information of adjacent codons \cite{Liu}. In particular, it has been shown that there is a ``strong short-range correlation of adjacent bases'' and that adjacent genes are more likely to be co-expressed than non-adjacent genes \cite{Luo} \cite{Cohen}. Additionally, the expression patterns of adjacent genes are more correlated than patterns of randomly selected gene pairs \cite{Kruglyak}. In all, various studies have demonstrated that the nearby informational units have an effect on the current informational unit \cite{Liu}. 

This is important, yet has not been heavily factored into models; therefore, a convolutional code model that considers the effects of adjacent information is better for studying DNA encoding as opposed to a block code model that only considers the effects of the current information \cite{Liu}. 

2. Additionally, it is important to consider how informational units should be defined. In traditional methods, a single nucleotide in a DNA sequence is considered as an independent information unit, but understanding how to treat an entire codon as an information unit would be more representative of processes such as DNA translation \cite{Liu}.

3. The phenomenon of codon degeneracy, also known as the redundancy of the genetic code, should also be considered when creating models to represent biological processes. With four DNA bases, human bodies are able to develop 64 possible combinations of codons. Although there are 64 possible codons, there are only 20 different amino acids which these codons represent. Thus, various codons can correlate to more than one amino acid. This degeneracy of codons provides more stability in genetic processes, such that a nonsense mutation occurs in which ``a gene mutation of one nucleotide may result in another codon of the same amino acid'' \cite{Liu}



Ultimately, the use of a convolutional code model views the encoding object, say a ribosome, as a mechanism with memory \cite{May-cc}. As convolutional coding produces encoded blocks based on present and past bits of information, "the modeling assumption is that genetic operations such as initiation and translation may involve 'decisions' which are based on immediate past and immediate future information" \cite{Elebeoba}. Presuming mRNA to be convolutionally encoded data allows the convolutional code model to factor in the inter-relatedness that is present between bases in an mRNA sequence \cite{Elebeoba}.

\subsection{Current Convolutional Code Models for DNA and Biological Phenomena}


\textbf{Liu and Geng:} In a study published in 2013, Liu and Geng devised a novel approach to encode DNA sequences through a $(6, 3, 2)$ convolutional code model that takes into account factors 1, 2, 3 above and is species-independent \cite{Liu}. This study treated a codon as an information unit and had a generator matrix based on the principle of codon degeneracy. 

As the codon is treated as an information unit, 3 or multiples of 3 are defined as the basic code length. Due to the short-range dominance of base correlation, the universal constraint length $L$ is set to $2$ \cite{McGraw-Hill}. Liu and Geng defined the length of the convolutional code as 6 \cite{Liu}. This all implies that the encoder output depends on 2 adjacent codons and the code is a $(6, 3, 2)$ convolutional code such that the inputs are sets of 3 nucleotides grouped together with 6 output symbols. 

The coefficients in the generator matrix are based on the principle of codon degeneracy, meaning that the resultant amino acid for the codons could be the same for multiple codons. The wobble effect was taken into consideration such that the first two bases of each group of 3 input sequences were given a higher weight while the third base was given a lower weight as it is affected by the wobble effect. 

The generator matrix for the $(6, 3, 2)$ convolutional code was defined such that $g_{k, n}^l$ denotes whether, at time $t,$ the input data of the $k-th$ row $(k = 1, 2, 3)$ and the $l-th$ column $(l = 1, ...  , 6)$ acts on the output $C_{n}^i ( n = 1, ... , 6)$. When $g_{k, n}^l = 1$, there is a solid line between the message and the binary addition operator in the shift register of the code. When $g_{k, n}^l$ = 0, there is no solid line or no direct influence.  

The encoder outputs $ $C$ = 
\begin{bmatrix} 
m^2g^1 + m^1g^2
\end{bmatrix},
$
where $m^2$ is the current input block of length 3 and $m^2$ is the previous block of length 3.  

So the generator matrix for the code is $g^1 = g^2 =  
\begin{bmatrix}
 110110\\
 110110\\
110110\\
\end{bmatrix}
$ 

  
  
The efficacy of this code was tested by taking the DNA sequences of 12 prokaryote and 9 eukaryote model organisms that contained differing GC content. Notably, this paper digitized the nucleotide sequences by assigning 0, 1, 2, 3 respectively to A, G, C, T. This differs from what Nemzer proposed and the DNA sequences used for analysis can be digitized differently in further studies based on Nemzer's proposition. After the sequence was digitized, the output was calculated based on the specific convolutional code, and the Hamming distance was calculated between the first three bits of the  output and the previous input data. The open reading frames (ORFs) and termination site DNA sequences were analyzed for all the DNA sequences and the Hamming distances were obtained. The characteristic average code distances (CACD) of the DNA sequence were calculated. 

The results of this study indicate that the proposed convolutional code model can be used as a taxonomic characteristic, based on CACD and GC content, for studying the relationships of organisms. More research should be conducted on understanding the effects of outside influences like horizontal gene transfer or biotechnology techniques. Additionally, it is vital to better understand the relationship between CACD and GC content.


\textbf{Press et al.:} In August 2020, a study published by Press et al. proposed the construction of an error-correcting code for DNA storage that was able to correct insertion, deletion, and substitution mutations while adhering to a class of user-defined constraints \cite{Press}. A tested algorithm was proposed that applied convolutional and block error-correcting codes for recovering error-free DNA data. When given message bits, the HEDGES algorithm, also known as the Hash Encoded, Decoded by Greedy Exhaustive Search, emits DNA characters. 

While DNA codes have a code rate of r = 1.0 and thus no error correction is possible, it is vital to introduce efficient error-correcting codes (ECC) that can recover the original information \cite{Press}. Insertions and deletions represent 50$\%$ of observed DNA errors, yet most DNA encoding schemes consist of ECCs that correct substitutions. Previous DNA storage implementations correct insertions and deletions (indels) through high depth sequencing, multiple alignment, and consensus base calling, yet this is inefficient repetition. The HEDGES algorithm presented a method to correct indels and substitutions with  high code rates and minimum redundancy in stored DNA \cite{Press}. 

There are two parts to the HEDGES algorithm: the inner code and the outer code. The inner code, also termed HEDGES, translates from the DNA code alphabet {A, C, T, G} and the binary representation {0, 1}. It also corrects most of the indels and substitution errors while adhering to various constraints. The HEDGES inner code is a convolutional code (tree code) with an infinite length, as the code reflects properties that are specific to DNA. The code is decoded through a stack algorithm. The outer code, which is a Reed-Solomon code, corrects any residual errors with a high probability. 

Error correction is a major constraint in biology, especially when considering the extent to which DNA repair can effectively identify mutations and then fix them. Homopolymer runs, imbalanced GC content, and dropout errors, in addition to the general chromosomal and base-pair mutations, are very common and problematic. The HEDGES algorithm adeptly proposes a code that corrects insertions, deletions, and substitutions. The algorithm was tested in silico and with synthesized DNA strands from which a statistical model for larger datasets was developed. 

 
\textbf{May et al.:} In various experiments done with ribosomes by May et al., the translation initiation process of messenger RNA sequences is analyzed and a convolutional code mdoel is proposed. May et al. believed that the translation of mRNA could be viewed as the decoding of information sequences and that mRNA sequences could be viewed as noisy convolutionally encoded signals \cite{May-cc}. They initially proposed that convolutional codes could effectively model genetic processes like initiation and translation as both take present and past information into consideration. 

May et al. used table-driven encoders and decoders, and the methods for encoding and decoding are described \cite{May-cc}. In experiments, ribosomes are functionally parallel to a table-based convolutional decoder, and the 16s rRNA sequence is used to form decoding masks for the decoding \cite{May-cc}. In all, the research conducted by May et al. proved the potential of convolutional code based models for representing genetic operations, especially those that could distinguish between translated (protein coding) and non-translated (non-protein coding) genomic sequences. 

\textbf{Ponnala et al.:} Building on the findings of May et al., Ponnala et al proposed convolutional code generators, in which the 16s rRNA was used, for translation initiation in E. coli K-12 \cite{Ponnala Thesis} \cite{Ponnala}. The ribosome was again modeled as table-based convolutional decoders such that they were also viewed as noisy, convolutionally encoded signals similar to the study by May et al. Based on the ability to produce encoded sequences that clearly distinguished between translated and non-translated sequences, generators were chosen. Table-based convolutional coding was applied to find the generators, and the methodology for doing so was provided by Ponnala et al. \cite{Ponnala}.

\subsection{Proposal of New Convolutional Code Model for DNA}

This study proposes an adopted version of the convolutional code model built by Liu and Geng \cite{Liu}. They determined the coefficients for the generator matrix based on codon degeneracy and the wobble effect: the first two bases of each codon block are assigned a higher weight while the third base is given a lower weight. While this is a fairly accurate representation, it can be improved by implementing a hierarchical weight distribution system such that the weights of the bases are ordered such that $w_{b_2} > w_{b_1} > w_{b_3}$. Various studies have confirmed the fact that the second base in a codon is the most important as it conveys the most information and specifies the specific type of amino acid \cite{Saier}. The first base should be second in weight as it determines the amino acid, and position 3 should be last in weight due to the wobble effect. 

Additionally, in the shift register diagram that Liu and Geng have proposed, the third equation is assigned 0 message bits, implying that it has no contribution to the final encoded message. This should be adjusted such that the third equation should take into account the first two bases of each of the codons such that it corresponds to four bases total in a block of six bases. 

By incorporating these changes and adjusting the weight, a more accurate convolutional code model will be developed. These adjustments are in progress and will be tested. 



\section{Conclusion}


Convolutional codes are  linear codes that take into consideration past and present information bits. They are effective when used in models, algorithms, and decoding schemes for DNA codes due to their ability to incorporate genetic phenomena like codon context. Various studies conducted by Liu and Geng, Press et al., May et al., and Ponnala et al. examine the role of convolutional codes in biological phenomena. This study explains fundamental properties of convolutional and DNA codes, examines the applications of convolutional codes to DNA codes, and proposes revisions for one currently existing model. These revisions are currently being incorporated and will be tested to assess their efficiency. Future extensions of this paper also include using convolutional neural network (CNN) based techniques to identify and correct errors in DNA sequences. These adjustments demonstrate the ever-growing field of biocomputation. While the applications of convolutional codes to DNA codes is one sub-field of biocomputation, this field will continue to expand by building on the already established knowledge.

\pagebreak

\section{References}
 
 \renewcommand{\section}[2]{}
 
\end{document}